\documentclass[utf8]{FrontiersinHarvard} 

\usepackage{url,hyperref,lineno,microtype}
\usepackage[onehalfspacing]{setspace}

\def\keyFont{\fontsize{8}{11}\helveticabold }
\def\firstAuthorLast{Civano {et~al.}} 
\def\Authors{F.~Civano\,$^{1,*}$, X.~Zhao\,$^{2}$, P.~G.~Boorman\,$^{3}$, S.~Marchesi\,$^{4,5}$, T.~Ananna\,$^{6}$, S.~Creech\,$^{7}$, C-T.~Chen\,$^{8,9}$, R.~C.~Hickox\,$^{10}$, D.~Stern\,$^{11}$, K.~Madsen\,$^{1}$, J.~A.~García\,$^{1}$, R.~Silver\,$^{1}$, J.~Aird\,$^{12}$,
D.~M.~Alexander\,$^{13}$,
M.~Balokovi\'{c}\,$^{14,15}$,
W.~N.~Brandt\,$^{16,17,18}$,
J.~Buchner\,$^{19}$,
P.~Gandhi\,$^{20}$,
E.~Kammoun\,$^{21,22}$,
S.~LaMassa\,$^{23}$,
G.~Lanzuisi\,$^{4}$,
A. Merloni\,$^{19}$,
A. Moretti\,$^{24}$,
K. Nandra\,$^{19}$,
E.~Nardini\,$^{21}$,
A.~Pizzetti\,$^{5}$,
S.~Puccetti\,$^{25}$,
R.~W.~Pfeifle\,$^{1}$,
C.~Ricci\,$^{26,27}$,
D. Spiga\,$^{28}$,
N.~Torres-Alb\`{a}\,$^{5}$,
and the \textit{HEX-P} Collaboration
}
\def\Address{\footnotesize

$^{1}$NASA Goddard Space Flight Center, Greenbelt, MD 20771, USA \\
$^{2}$Harvard-Smithsonian Center for Astrophysics, 60 Garden Street, Cambridge, MA 02138, USA\\
$^{3}$Cahill Center for Astrophysics, California Institute of Technology, 1216 East California Boulevard, Pasadena, CA 91125, USA \\
$^{4}$INAF—Osservatorio di Astrofisica e Scienza dello Spazio di Bologna, Via Piero Gobetti, 93/3, I-40129, Bologna, Italy \\
$^{5}$Department of Physics and Astronomy, Clemson University, Kinard Lab of Physics, Clemson, SC 29634, USA \\
$^{6}$ Department of Physics and Astronomy, Wayne State University, 666 W Hancock St, Detroit, MI 48201, USA \\
$^{7}$Department of Physics \& Astronomy, University of Utah, 115 South 1400 East, Salt Lake City, UT 84112, USA \\
$^{8}$ Science and Technology Institute, Universities Space Research Association, Huntsville, AL 35805, USA\\
$^{9}$Astrophysics Office, NASA Marshall Space Flight Center, ST12, Huntsville, AL 35812, USA\\
$^{10}$Department of Physics and Astronomy, Dartmouth College, Hanover, NH 03755 \\
$^{11}$Jet Propulsion Laboratory, California Institute of Technology, Pasadena, CA 91109, USA \\
$^{12}$Institute for Astronomy, University of Edinburgh, Royal Observatory, Edinburgh EH9 3HJ, UK\\
$^{13}$Centre for Extragalactic Astronomy, Department of Physics, Durham University, UK \\
$^{14}$Yale Center for Astronomy \& Astrophysics, 52 Hillhouse Avenue, New Haven, CT 06511, USA \\
$^{15}$Department of Physics, Yale University, P.O. Box 2018120, New Haven, CT 06520, USA \\
$^{16}$Department of Astronomy \& Astrophysics, The Pennsylvania State University, University Park, PA 16802, USA \\
$^{17}$Institute for Gravitation and the Cosmos, The Pennsylvania State University, University Park, PA 16802, USA \\
$^{18}$Department of Physics, The Pennsylvania State University, University Park, PA 16802, USA \\
$^{19}$Max-Planck-Institut f\"{u}r extraterrestrische Physik, Giessenbachstra{\ss}e 1, D-85748 Garching, Germany \\
$^{20}$Department of Physics \& Astronomy, Faculty of Physical Sciences and Engineering, University of Southampton, UK \\
$^{21}$INAF -- Osservatorio Astrofisico di Arcetri, Largo Enrico Fermi 5, 50125 Firenze, Italy \\
$^{22}$Dipartimento di Matematica e Fisica, Universit\`{a} Roma Tre, 00146 Rome, Italy\\
$^{23}$Space Telescope Science Institute, 3700 San Martin Drive, Baltimore, MD 21210, USA \\
$^{24}$INAF, Osservatorio Astronomico di Brera, Via Brera 28, 20121 Milano, Italy\\
$^{25}$ASI - Agenzia Spaziale Italiana, Via del Politecnico snc, 00133 Roma, Italy \\
$^{26}$N\'{u}cleo de Astronom \'{i}a de la Facultad de Ingenier\'{i}a, Universidad Diego Portales, Av. Ej\'{e}rcito Libertador 441, Santiago, Chile \\
$^{27}$Kavli Institute for Astronomy and Astrophysics, Peking University, Beijing 100871, China \\
$^{28}$INAF, Osservatorio Astronomico di Brera, Via Bianchi 46, 23807 Merate, Italy}
\def\corrAuthor{Francesca Civano}

\def\corrEmail{francesca.m.civano@nasa.gov}
\def\sixte{\texttt{SIXTE}}

\def\xmm{{XMM-{\it Newton\/}}}
\def\nustar{{\it NuSTAR}}

\def\cgs{{ erg s$^{-1}$} cm$^{-2}$}
\def\arcsec{{''}}

\newcommand{\chandra}{\textit{Chandra\/}}

\begin{document}
\onecolumn
\firstpage{1}

\title {The High Energy X-ray Probe (HEX-P): Bringing the Cosmic X-ray Background into Focus}

\author[\firstAuthorLast ]{\Authors} 
\address{} 
\correspondance{} 

\extraAuth{}

\maketitle

\begin{abstract}

Since the discovery of the cosmic X-ray background (CXB), astronomers have strived to understand the accreting supermassive black holes (SMBHs) contributing to its peak in the 10--40 keV band. Existing soft X-ray telescopes could study this population up to only 10 keV, and, while \nustar\ (focusing on 3--24 keV) made great progress, it also left significant uncertainties in characterizing the hard X-ray population, crucial for calibrating current population synthesis models. This paper presents an in-depth analysis of simulations of two extragalactic surveys (deep and wide) with the \textit{High-Energy X-ray Probe} (\textit{HEX-P}), each observed for 2 Ms. Applying established source detection techniques, we show that \textit{HEX-P} surveys will reach a flux of $\sim$10$^{-15}$ \cgs\ in the 10--40 keV band, an order of magnitude fainter than current \nustar\ surveys. With the large sample of new hard X-ray detected sources ($\sim2000$), we showcase \textit{HEX-P}'s ability to resolve more than 80\% of the CXB up to 40 keV into individual sources. The expected precision of \textit{HEX-P}'s resolved background measurement will allow us to distinguish between population synthesis models of SMBH growth. \textit{HEX-P} leverages accurate broadband (0.5--40 keV) spectral analysis and the combination of soft and hard X-ray colors to provide obscuration constraints even for the fainter sources, with the overall objective of measuring the Compton-thick fraction. With unprecedented sensitivity in the 10--40 keV band, \textit{HEX-P} will explore the hard X-ray emission from AGN to flux limits never reached before, thus expanding the parameter space for serendipitous discoveries. Consequently, it is plausible that new models will be needed to capture the population \textit{HEX-P} will unveil.


\tiny
 \keyFont{ \section{Keywords:} X-ray, surveys, AGN, obscuration, models, cosmic X-ray background, Compton-thick} 
\end{abstract}


\section{Introduction}\label{sec:intro}

The cosmic X-ray background (CXB), i.e., the diffuse emission from 0.1--100 keV discovered in the early 1960s by \cite{giacconi62}, has been a focal point at the center of X-ray astronomy research since then. As most of the emission between 10--100 keV is expected to come from accretion of matter onto supermassive black holes (SMBHs; see \citealt{ueda99}, \citealt{Gandhi03}, \citealt{gilli07}, \citealt{Comastri15}, \citealt{ananna19}), the CXB provides a crucial verification for current models of SMBH growth across cosmic time.
The growth of these SMBHs, which are known as active galactic nuclei (AGN) during phases of accretion, is intricately linked to the evolution of galaxies and their episodes of star formation (see \citealt{harrison2017}). Consequently, studying the population contributing to the CXB offers invaluable insights into the co-evolution of SMBHs and galaxies.

We now understand that the CXB represents the total integrated emission of faint X-ray sources across the sky. Thanks to the extragalactic surveys (e.g., \citealp{civano16},\citealp{luo2017}, \citealp{chen2018}) performed over the past $\sim$25 years, particularly with the use of focusing soft X-ray telescopes like \chandra\ and \xmm, we have been able to resolve and study the accreting SMBHs that contribute to approximately $\sim$80--85\% of the CXB at energies below 10 keV \citep[e.g.,][]{wors05,Hickox06,moretti03,moretti12,Cappelluti17,xue12stack}. The large samples of AGN in these soft band surveys have allowed us to constrain various properties including, e.g., the evolution of the X-ray luminosity functions of AGN out to $z\sim5$ \citep{vito14,Buchner15,aird15} and the evolution of the unobscured ($N_{\rm H} < 10^{22}$ cm$^{-2}$) and obscured ($N_{\rm H} > 10^{22}$ cm$^{-2}$) fraction to comparable redshifts \citep[e.g.,][]{Marchesi16,vito18,peca23}.  

Compton-thick AGN (CT; $N_{\rm H} > 10^{24}$ cm$^{-2}$) are a crucial piece of the CXB that still remains unresolved. Soft X-ray surveys have faced challenges to detect CT AGN as the low--energy photons ($<$10 keV) are easily absorbed by heavily obscuring column densities.  In particular, at low redshifts, most of the emission from CT AGN can only be observed at energies greater than 10 keV, making softer energy X-ray surveys less suitable to detect these populations. In contrast, hard X-ray photons ($>$10 keV) can penetrate higher column densities. High-redshift ($z>1$--2) CT sources can be detected in soft surveys due to their main signature, the Compton-hump (20--30 keV), moving to $<$10 keV. However, collecting large samples requires deep exposures at very faint fluxes, limiting observations to only a few fields and introducing significant uncertainties on the fraction of CT AGN \citep[e.g.,][]{Buchner14, Brightman14,lanzuisi18,Yan23}. 


The launch of \nustar\ \citep{Harrison13} and the extragalactic surveys that were performed in the hard X-rays (3--24 keV) 
have provided a breakthrough, enabling us to observe the high-energy X-ray universe where obscuration is less of an issue, and CT sources can be more easily detected. \nustar\ adopted the classic ``wedding cake'' survey strategy by observing the most notable extragalactic fields \citep[COSMOS, ECDFS, EGS, CDFN and UDS;][]{civano15,mullaney15,masini18}, and a survey of all the serendipitous sources in \nustar\ observations \citep{Alexander13,Lansbury17}. Moreover, \nustar\ is currently observing the deepest survey ever performed in the 3--24 keV range in the North Ecliptic Pole, a \textit{JWST} Time Domain Field, reaching fluxes of 3.7$\times$10$^{-14}$ \cgs\ (over 50\% of the area) in the 8--24 keV band \citep{zhao21,zhao24}. These \nustar\ surveys have probed the demographics of hard X-ray emitting AGN up to $z\sim3$, providing the most precise measurements of the hard X-ray flux distribution ($\log N-\log S$) and luminosity function \citep{Harrison16,Aird15_nustar,zhao24} and directly resolving 35\% of the CXB in the 8--24 keV range with detected sources \citep{Harrison16,Hickox24}.
Interestingly, \nustar\ surveys have shown that previously detected X-ray AGN are significantly more obscured than can be determined through soft X-rays alone and even CT \citep{civano15}, and that luminous obscured AGN selected at other wavelengths (e.g., {\em WISE}) remain extremely weak or undetected in deep \nustar\ exposures, implying $N_{\rm H} > 10^{25}$ cm$^{-2}$ \citep{Stern14,Yan19,carroll23}. 
Spectral analysis of the 63 brightest hard X-ray sources in \nustar\ surveys \citep{zap18} has shown that broadband data covering the 0.5--24 keV band (\chandra\ or \xmm\ combined with \nustar) are essential for constraining the spectral parameters such as photon index, column density, reflection and luminosity. This is in agreement with results at $z=0$ where \citet{marchesi17,marchesi18,marchesi19} have shown that the lack of sensitive hard X-ray coverage results in an overestimate of the CT-AGN fraction. While \nustar\ has made substantial progress in measuring the spectral properties of CT AGN in the local Universe, there are still significant inconsistencies between the observed and predicted CT fraction \citep[see][and references therein]{boorman23}. At higher redshift, the CT fraction remains unconstrained, with an upper limit of 66\% at 90\% confidence level \citep{zap18}. Similarly, results solely from hardness ratio analysis have led to very large uncertainties on the CT fraction at $z>0.1$ \citep{civano15,masini18,zhao21,zhao24}.

Population synthesis models \citep[e.g.,][]{comastri95,ballantyne06,gilli07,treister09,ueda14,ananna19} combine AGN spectral shapes and AGN space densities in luminosity, obscuration, and redshift bins to reproduce the intensity and shape of the CXB. The space densities are constrained using the results from existing X-ray surveys described above. The completeness and robustness of such results are limited by the energy band probed (most of the sources are only detected by \chandra\ and \xmm\ below 10 keV) and by the survey's flux limits. AGN samples from higher energy surveys such as \nustar\ are quite scarce, and most of the hard X-ray constraints are from \textit{Swift}-BAT, therefore limited to bright, low redshift sources \citep{ricciswift2017,Oh_2018}. The typical AGN spectrum is another piece of the puzzle when calibrating population synthesis models: without high-energy X-ray data, constraining spectral parameters such as the reflection scaling factor (relevant at 20--40 keV) and the high-energy cutoff (relevant at $E$ $>$ 60 keV) remains challenging \citep{ricciswift2017,ananna2020a,Kammoun23}. As the dominant AGN contribution to the CXB is expected to be in the 10--100 keV energy band, poorly constrained spectral parameters can hinder the accuracy of population synthesis models. To justify the low resolved fraction of the CXB at high energy, the latest models have either changed assumptions on AGN spectral shapes \citep[e.g.,][]{Akylas12} or predicted that a large fraction of the sources contributing to the CXB must be extremely obscured or CT, and that the fraction should reach $\sim$50\% at faint fluxes \citep[$<$10$^{-15}$ \cgs\ at 20--30 keV;][]{ananna19}. Similar results are obtained in cosmological hydrodynamical simulations \citep{Ni22}. Recent observational evidence of an extremely obscured and possibly CT AGN at $z=10.1$ detected behind the Abell 2644 cluster \citep{bogdan23,goulding23} supports these models, implying a large CT fraction at high redshift. Moreover, results from {\it JWST} indicating a steepening at the faint end of the AGN luminosity function \citep{Harikane23, Maiolino23b, Maiolino23a} 
could suggest that, as we are finding more faint unobscured AGN than expected, the intensity of the CXB could be a factor of 10 higher than what was constrained before \citep{Padmanabhan23}. 

Resolving the population contributing to the CXB in the 10--40 keV band, determining their spectral properties and constraining the CT fraction at faint fluxes with the goal of providing strong constraints on population synthesis models are the major goals of the \textit{High-Energy X-ray Probe} \citep[\textit{HEX-P};][]{Madsen23}. In this paper, we present the results obtained from accurate simulations (carried out using the observatory performance based on current best estimates as of Spring 2023) of a deep narrow ($\sim$0.16 deg$^2$) and a shallow wide  ($\sim$1.1 deg$^2$) extragalactic survey, employing 2 Ms of exposure time each. We assume such surveys will be performed in well-known extragalactic fields, providing rich datasets (e.g. including data from Rubin, \textit{JWST}, \textit{Roman}) for a multiwavelength characterization of \textit{HEX-P} detections. 

The paper is structured as follows: Section 2 describes the telescope's characteristics; Section 3 explains the details of the survey simulations; Section 4 presents the source properties; Section 5 discusses the findings regarding the CXB resolved fraction; Section 6 presents the results related to obscured sources detected in the surveys, spectral analysis, and the CT fraction; and Section 7 includes a concise summary.

\section{The High Energy X-ray Probe}\label{sec:HEX-P}

\textit{HEX-P} is a probe-class mission concept that offers sensitive broad-band coverage (0.2--80\,keV) of the X-ray spectrum with exceptional spectral, timing and angular capabilities. It features two high-energy telescopes (hereafter HET; 2--80 keV) that focus hard X-rays, and a low-energy telescope (hereafter LET; 0.2--25 keV) that has soft X-ray coverage.

The LET consists of a segmented mirror assembly coated with Ir on monocrystalline silicon that achieves a half-power diameter of 3.5'', and a low-energy DEPFET detector of the same type as the Wide Field Imager \citep[WFI;][]{Meidinger20} onboard Athena \citep{nandra13}. It has 512 $\times$ 512 pixels that cover a field of view of 11.3’ $\times$ 11.3’. It has an effective passband of 0.2--25\,keV, and a full frame readout time of 2\,ms, which can be operated in a 128 and 64-channel window mode for higher count-rates to mitigate pile-up and faster readout. Pile-up effects remain below an acceptable limit of $\sim 1\% $ for a flux up to $\sim 100$\,mCrab (2--10 keV) in the smallest window configuration. Excising the core of the point spread function (PSF), a common practice in X-ray astronomy, will allow for observations of brighter sources, with a  typical loss of up to $\sim 60\%$ of the total photon counts.

The HET consists of two co-aligned telescopes and detector modules. The optics are made of Ni-electroformed full shell mirror substrates, leveraging the heritage of \xmm\ \citep{Jansen2001}, and coated with Pt/C and W/Si multilayers for an effective passband of 2--80\,keV. The high-energy detectors are of the same type as flown on \nustar\ \citep{Harrison2013}, and they consist of 16 CZT sensors per focal plane, tiled 4 x 4, for a total of 128 $\times$ 128 pixels spanning a field of view of 13.4’$\times$13.4’, slightly larger than that of the LET.

The broad X-ray passband of \textit{HEX-P} and superior sensitivity to \nustar\ will provide a unique opportunity to probe the evolution of AGN in extragalactic surveys over a wide range of obscuration regimes that is not possible with soft X-ray instruments alone. The spectral constraints provided by simultaneous coverage of the soft and hard X-ray bands will remove any issues associated with variability and provide spectral constraints for new sources detected above 10 keV. By resolving a high fraction of the CXB in hard X-rays, \textit{HEX-P} will ultimately enable far more sensitive population synthesis of SMBH growth in the context of the evolution of their host galaxies as probed by a plethora of multiwavelength observations.


\section{\textit{HEX-P} survey simulations}\label{sec:simulations}
The approach we used to perform end-to-end simulations of surveys with \textit{HEX-P} is essentially identical to the one adopted in \citet{marchesi20}, 
to which we refer the reader for a detailed explanation while providing an overview here. The analysis of our simulated data was developed based on \nustar\ extragalactic surveys \citep{civano15, mullaney15, masini18, zhao21}. Here, we briefly present the tool we used to perform the simulations, as well as the input mock catalogs of AGN and non-active galaxies and the analysis of such simulated data.

\subsection{The SIXTE simulation tool}
The \textit{HEX-P} surveys presented in this paper have been simulated using the Monte Carlo code Simulation of X-ray Telescopes \citep[hereafter \sixte,][]{dauser19}. This software enables the simulation of observations using an X-ray telescope in the following manner.
Initially, the tool creates a photon list, which includes the arrival time, energy and position of each photon. To generate this initial information, \sixte\ reads the instrument configuration from an {\it xml} file. This first step uses the instrument's effective area, the field of view and pointing. The photon list created in the first step is then convolved with the instrument PSF and vignetting. In doing so, \sixte\ generates an impact list that contains the energy and arrival time of each photon, as well as its position on the detector. The final event file is obtained from this intermediate list and reprocessed to take into account the simulated detector read-out properties and redistribution matrix file. 

The \textit{HEX-P} simulations presented in this work were produced with a set of response files that represent the observatory performance based on current best estimates as of Spring 2023 (v07; see \citealt{Madsen23}). The effective area is derived from ray-tracing the mirror design, including obscuration by all known structures. The detector responses are based on simulations performed by the hardware groups, with an optical blocking filter for the LET and a Be window and thermal insulation for the HET. The LET background was derived from a GEANT4 simulation \citep{Eraerds2021} of the WFI instrument. The HET background was derived from a GEANT4 simulation of the \nustar\ instrument. Both simulations adopt the planned L1 orbit for \textit{HEX-P}. The \sixte\ team has included the above \textit{HEX-P} configuration files (i.e., telescope setup, response matrices, vignetting, point spread function) for both the HET and LET in their system, ready to be used.

\subsection{Active galactic nuclei mock catalog}
In order to perform the steps above and to produce a realistic representation of the X-ray sky, \sixte\ needs an input source list generated in the \texttt{SIMPUT} data format. In this paper, we use the mock catalogs of AGN and non-active galaxies presented in \citet{marchesi20}. The catalogs we used are available \href{http://cxb.oas.inaf.it/mock.html}{online} in FITS format and ready to be used within \sixte. 

The mocks have been calibrated to reproduce known trends between AGN number densities and luminosity, redshift and column density. In detail, each AGN in the catalog has intrinsic 0.5--2\,keV luminosities, redshift and column density, computed by resampling the X-ray luminosity function of unabsorbed AGN given by \citet{hasinger05}, scaled up by a luminosity--dependent factor to account for the whole AGN population \citep[see][]{gilli07}. 
No assumption is made on AGN or host clustering. AGN have been simulated down to a 0.5--2\,keV luminosity $L_{0.5-2}$=10$^{40}$\,erg\,s$^{-1}$ and up to $z=10$. The mock AGN number counts match the observed ones over the whole range of fluxes sampled by current X-ray surveys. The mock AGN sample average CT fraction is $f_{\rm CT}$=39\,\%. As a reference, this is a similar figure to that found in the current AGN population synthesis models: \citet[][$f_{\rm CT}$=33\,\%]{ueda14}, \citet[][$f_{\rm CT}$=38\,\%]{Buchner15}, and \citet[][$f_{\rm CT}$=50\,\%]{ananna19}. These are all average values; however, all models adopt a luminosity-dependent CT fraction, consistent with what is observed in X-ray surveys of AGN \citep[see, e.g.,][]{ricci15,marchesi16a}.

Finally, while the \citet{gilli07} model is generally in close agreement with the observational results from X-ray surveys, in the high-redshift regime (i.e., at $z>3$, where the AGN space density starts declining) the discrepancy becomes more prominent, with the model underestimating the expected number of sources with respect to the observational evidence. For this reason, \citet{marchesi20} developed a separate $z>3$ catalog using as a reference the \citet{vito14} $z>3$ AGN luminosity function, which describes the observational evidence from the deepest X-ray surveys currently available \citep[e.g.,][]{vito18}.

In the simulations of \textit{HEX-P} surveys presented in this work, we also include the non-active galaxies mock catalogs presented in \citet{marchesi20} and derived from the peakM and peakG model $\log N-\log S$ by \citet{ranalli05}. The galaxies in the mock have X-ray flux information but lack redshifts and luminosity, so we do not include them in our source detection analysis. The presence of the galaxies in the simulation, however, ensures that the overall simulated emission closely matches the CXB intensity.

\subsection{\textit{HEX-P} surveys layout and tiling strategy}
We simulated two \textit{HEX-P} fields, a Deep and a Wide survey. Each survey has been simulated both with the LET camera and with the two HET cameras. The two HET cameras are then co-added. The total exposure for each survey is 2\,Ms. Since the LET field of view ($\sim$11.3$^\prime$ $\times$11.3$^\prime$) is slightly smaller than the HET field of view ($\sim$13.4$^\prime$$\times$$13.4^\prime$), we used two slightly different tiling strategies for the Wide and Deep surveys to ensure uniform coverage at low energies.

\begin{figure}
\centering
\includegraphics[width=0.45\textwidth]{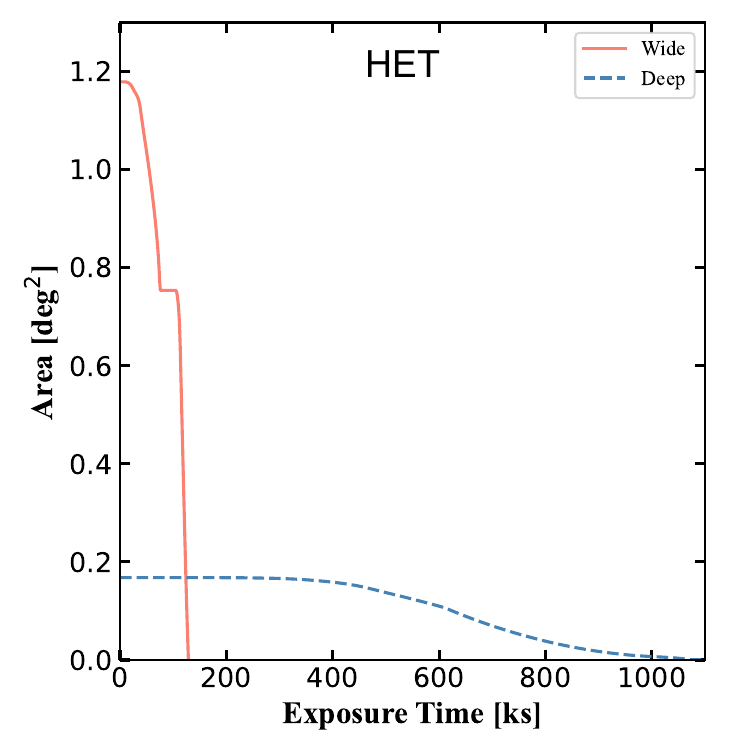}
\includegraphics[width=0.45\textwidth]{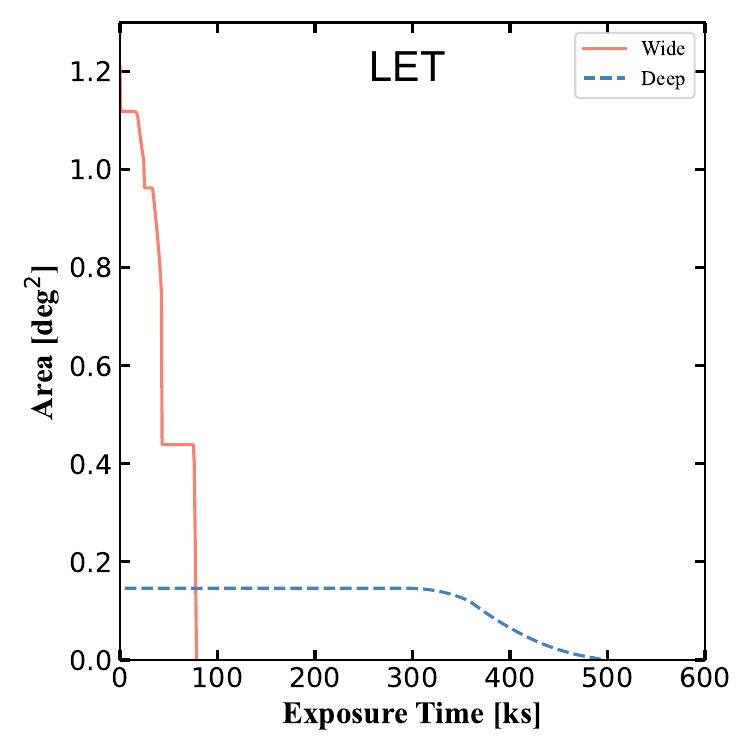}

\caption{\label{areaexpo}
Survey area as a function of the effective, i.e. vignetting-corrected, exposure time for the HET (left) in the 3--24 keV band and LET (right) in the 0.5--2 keV band. The deep and wide surveys are in blue and salmon, respectively.}
\end{figure}

The \textit{HEX-P} Deep Field is obtained by combining four 500\,ks observations. The survey tiling is a 2$\times$2 grid where the pointing centers are offset by 11.3$^\prime$ (i.e., a whole LET field of view) in right ascension or declination. Because the LET is smaller in size than the HET, the HET pointings are slightly overlapping by about 2$^\prime$. The overall HET survey area is $A_{\rm deep}$ = 0.1668\,deg$^{2}$, while the overall LET area is 0.1469\,deg$^{2}$ (see blue curves in Figure \ref{areaexpo} and \ref{sensitivity}).
 The \textit{HEX-P} Wide Field survey consists of 81 overlapping pointings, each having nominal exposures of 25\,ks, in a 9$\times$9 grid. The survey tiling is done using the half-a-field shift strategy, which was successfully used in the \nustar\ COSMOS survey \citep{civano15}: the center of each pointing is offset by 6.7$^\prime$ (i.e., half a HET field of view) in right ascension or declination with respect to the previous one. The overall survey exposure per detector is therefore 25\,ks $\times$ 81 pointings = 2.025\,Ms. The overall HET survey area is $A_{\rm wide}$ = 1.083\,deg $\times$ 1.083\,deg = 1.174\,deg$^{2}$, while the overall LET area is  1.12\,deg$^{2}$ (see salmon curves in Figure \ref{areaexpo} and \ref{sensitivity}).
    
\begin{figure}
\centering
\includegraphics[width=\textwidth]{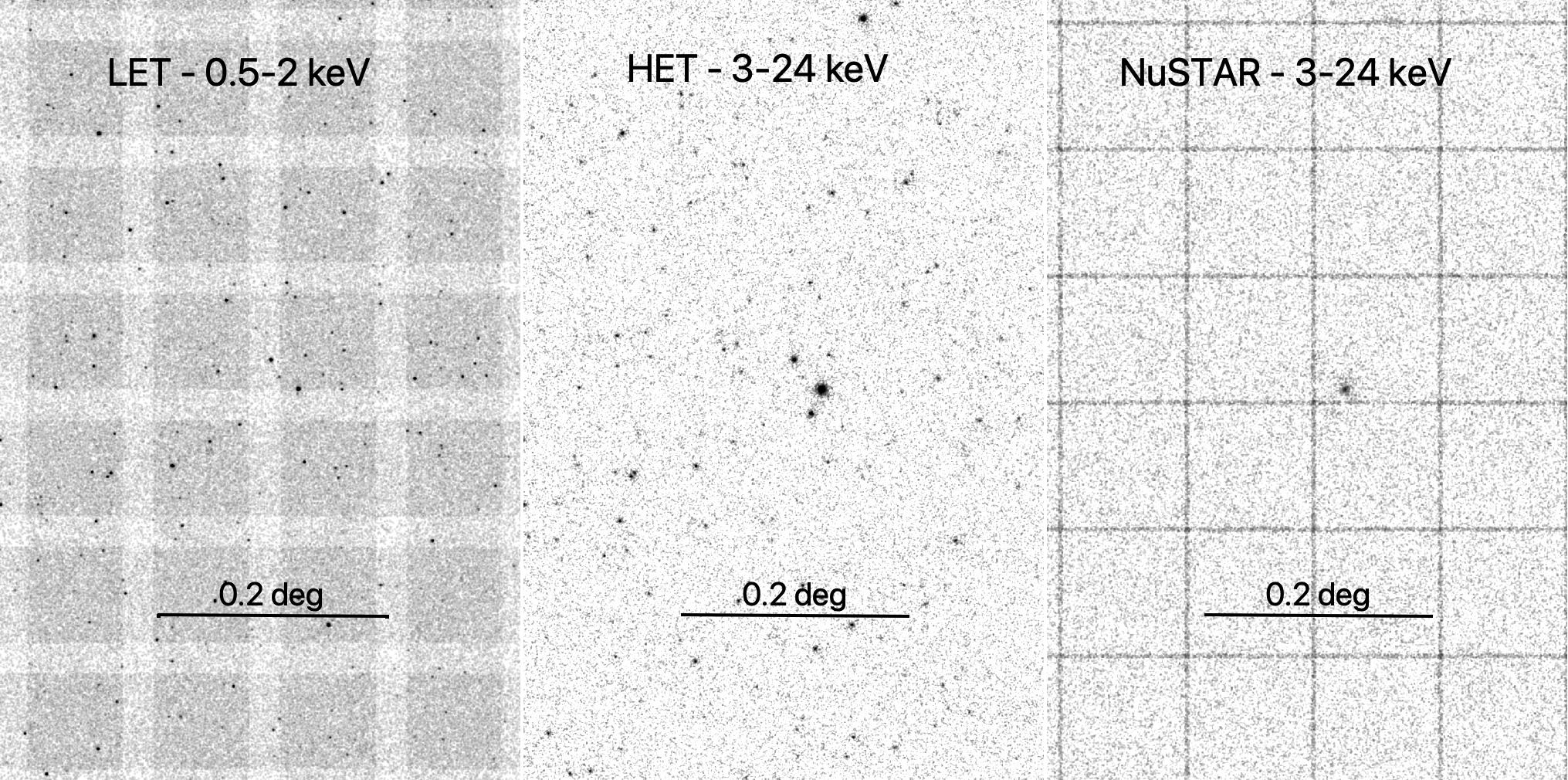}

\caption{\label{mosaics} A zoomed in section of the wide survey mosaics for the LET (0.5--2 keV), HET (2 telescopes summed, 3--24 keV) and \nustar\ (2 telescopes summed, 3--24 keV; this simulation was performed with \sixte). Hundreds of sources are visible in the HET and LET mosaics.}
\end{figure}

The area versus exposure for the deep (blue) and wide (salmon) surveys are shown in Figure \ref{areaexpo} for the HET (left) and the LET (right).
A zoomed in section of the wide survey mosaics for the LET (0.5--2 keV), HET (2 telescopes summed, 3--24 keV) and \nustar\ (2 telescopes summed, 3--24 keV) is shown in Figure \ref{mosaics}. The areas at lower exposure due to the overlapping of the tiles in the LET is visible but does not really affect the source detection (see Section 3.4). While hundreds of sources are visible by eye in the HET and LET images, only one source is clearly visible with \nustar\ using the same exposure.

\subsection{Source detection}
We performed source detection on the simulated data in five energy bands for the HET (3--8, 8--24, 3--24, 10--40 and 35--55 keV) and two bands for the LET (0.5--2 and 2--10 keV) following the procedure used in previous \nustar\ surveys  \citep{civano15, mullaney15, masini18, zhao21}. All the parameters used at each step (smoothing and matching radius, probability and significance thresholds, etc) were chosen to maximize the detection rate, the completeness and the reliability in the simulated \textit{HEX-P} surveys. We summarize the process in the following. 

First, we used \textit{Source-Extractor} \citep{Bertin1996} on the false-probability maps of the two surveys, which measures the probability ($P_{\rm false}$) that a signal is due to a background fluctuation rather than a real source. The false-probability maps were generated from the smoothed simulated maps (with a 7.5\arcsec\ and 3\arcsec\ smoothing radius for the HET and LET, respectively) and the background mosaics at each pixel using the incomplete Gamma function. The detection limit in \textit{Source-Extractor} is set to be $P_{\rm false}$ $\le$10$^{-p}$ (where $p$ is determined to maximize the detection rate). The coordinates of the detected sources were then used to extract the Poisson probability ($P_{\rm random}$) of each source, which is used to characterize the probability that a detection is due to a random fluctuation of the background. The Poisson probability is calculated using the total and background counts extracted from the simulated maps and the background maps at the coordinate of the detection using a circular aperture of 10\arcsec\ and 5\arcsec\ radius for the HET and LET, respectively. We then define the maximum likelihood (DET\underline{\;\;}ML) of each detection as the inverse logarithm of the Poisson probability, i.e., DET\underline{\;\;}ML = --\,ln\,$P_{\rm random}$. Therefore, a lower $P_{\rm random}$ (i.e., a higher DET\underline{\;\;}ML) suggests a lower chance that the signal arises from a background fluctuation. Multiple detections of the same X-ray source ($\sim$2\%) were removed using a radius of 20\arcsec. The detections were then matched with the input mock catalog using 10\arcsec\ and 7\arcsec\ searching radii for the HET and LET catalogs, respectively. 

\subsection{Reliability, completeness and survey sensitivity}

To evaluate and maximize the accuracy and efficiency of the source detection in actual observations, we compute the survey's reliability and completeness. Reliability is the ratio of the number of detected sources matched to input sources to the total number of detected sources at or above a particular DET\underline{\;\;}ML. Completeness is defined as the ratio of the number of detected sources matched to the input catalog and above a chosen reliability threshold to the number of sources in the input catalog at a particular flux level. The definition of reliability and completeness can be found in \citet{zhao21}, and references therein. 

For the \textit{HEX-P} surveys presented in this paper, we choose to set a reliability of 99\%, which implies a 1\% spurious detection rate. Reliability and completeness depend on exposure, and if the observation tiling is planned in a way to have the field exposure uniform, it is possible to adopt a single DET\underline{\;\;}ML threshold. The DET\underline{\;\;}ML values are listed in Table~\ref{Table:number_of_sources} along with the number of sources detected and matched to the input catalogs above the threshold in each energy band.


\begingroup
\renewcommand*{\arraystretch}{1.5}
\begin{table*}
\caption{DET\underline{\;\;}ML at 99\% reliability in different energy bands. Number of detections. Sensitivities at 20\%-area and 50\%-area.}
\centering
\label{Table:number_of_sources}
\resizebox{\textwidth}{!}{  \begin{tabular}{l|cccccc|ccc}
       \hline
       \hline     
&3--8\,keV&8--24\,keV&3--24\,keV&10--40\,keV&35--55\,keV&Total&0.5--2\,keV&2--10\,keV&Total\\
	\hline
	& \multicolumn{6}{c|}{Wide} & &\\
	\hline
	DET\underline{\;\;}ML(99\%) threshold&4.04&3.12&2.64&3.12&2.62&&2.65&2.53\\
	N$\rm_{detection}$&892&771&1632&806&36&1702&2261&2816&3611\\
	Sensitivity (20\%-area) 10$^{-15}$ cgs&2.00&3.90&2.74&4.66&14.5&&0.22&0.58\\
	Sensitivity (50\%-area) 10$^{-15}$ cgs&3.2&6.25&4.63&7.62&27.8&&0.51&1.28\\
	\hline
	& \multicolumn{6}{c|}{Deep}  & & \\
	\hline
	DET\underline{\;\;}ML(99\%) threshold&3.75&1.0&4.11&3.77&3.16&&3.03&3.46\\
	N$\rm_{detection}$&476&434&521&387&36&608&859&940&1188\\
	Sensitivity (20\%-area) 10$^{-16}$ cgs&3.6&6.6&9.3&12.0&33.9&&0.47&1.43\\
	Sensitivity (50\%-area) 10$^{-16}$ cgs&7.1&11.5&17.1&21.7&65.4&&0.92&2.74\\
	\hline 
	\hline
\end{tabular}}
\end{table*}
\endgroup

\begin{figure}
\centering
\includegraphics[width=0.45\textwidth]{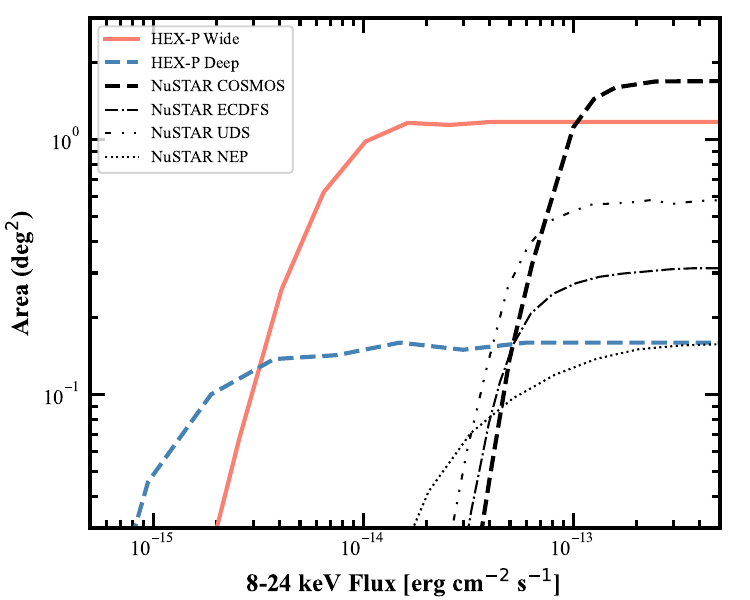}
\includegraphics[width=0.45\textwidth]{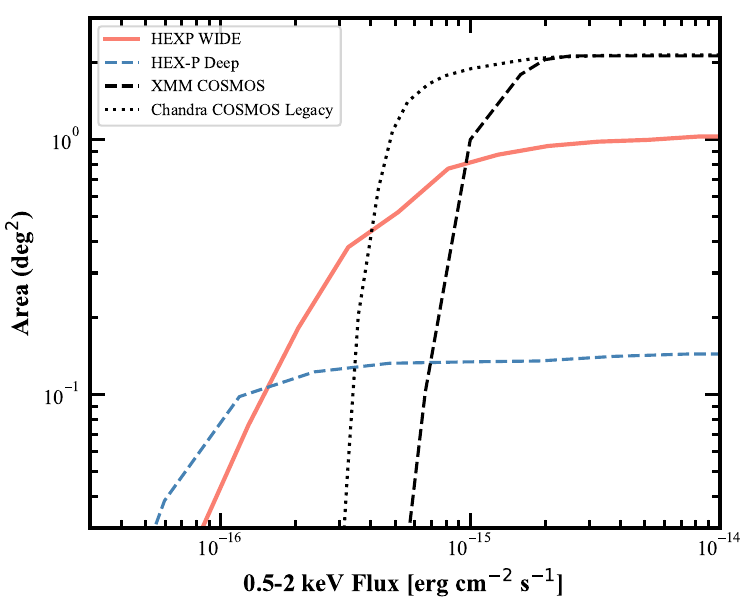}

\caption{\label{sensitivity}Survey sensitivity for the HET (left) in the 8--24 keV band and LET (right) in the 0.5--2 keV band. The deep and wide surveys are in blue and salmon, respectively. In each panel, the predicted sensitivities are compared with surveys published in the literature using \nustar\ in the left panel and \chandra\ and \xmm\ in the right panel. }
\end{figure}

The sky coverage (or the sensitivity) of the survey at a given flux can be derived from the completeness curve: if, at the chosen detection threshold, the completeness is sufficiently high (with reliability also high), the number of detected sources should correspond to the number of input sources with DET\underline{\;\;}ML greater than the threshold value. In this case, the sky coverage is the normalized version of the completeness curve to the total area of the survey. The sensitivity curves for the LET and HET wide and deep surveys are shown in Figure \ref{sensitivity} compared with current \nustar\ surveys for the HET and \chandra\ and \xmm\ surveys for the LET. Thanks to the large effective area, the lower background and the smaller PSF, \textit{HEX-P} surveys reach significantly fainter fluxes than any previous \nustar\ survey. The \textit{HEX-P} wide area survey is comparable in area and total exposure time (1.2 deg$^2$ and 2 Ms) to the COSMOS \nustar\ survey (1.7 deg$^2$ and 3 Ms) and reaches a $\sim$20 times fainter flux limit in the 8--24 keV band, as shown in Figure \ref{sensitivity}. The deep \textit{HEX-P} survey is comparable in area and exposure to the \nustar\ North Ecliptic Pole (\nustar-NEP; 0.16 deg$^2$ and 1.6 Ms) survey \citep{zhao21}, the deepest survey that \nustar\ has ever performed, but it is a factor of 25 times deeper in flux in the 8--24 keV band. In the lower energy regime, the wide LET survey is comparable in area covered with literature \chandra\ and \xmm\ COSMOS surveys \citep{civano16,cappelluti09}, but reach flux limits 4--5 times deeper employing roughly the same exposure time (e.g, \chandra\ COSMOS Legacy Survey used 4.6 Ms to cover 2.2 deg$^2$, XMM-COSMOS is slightly shallower and used 1.5 Ms to cover 2.2 deg$^2$, while the \textit{HEX-P} wide survey employs 2 Ms of time to cover 1.1 deg$^2$). 

\begin{figure}
\centering
\includegraphics[width=0.45\textwidth]{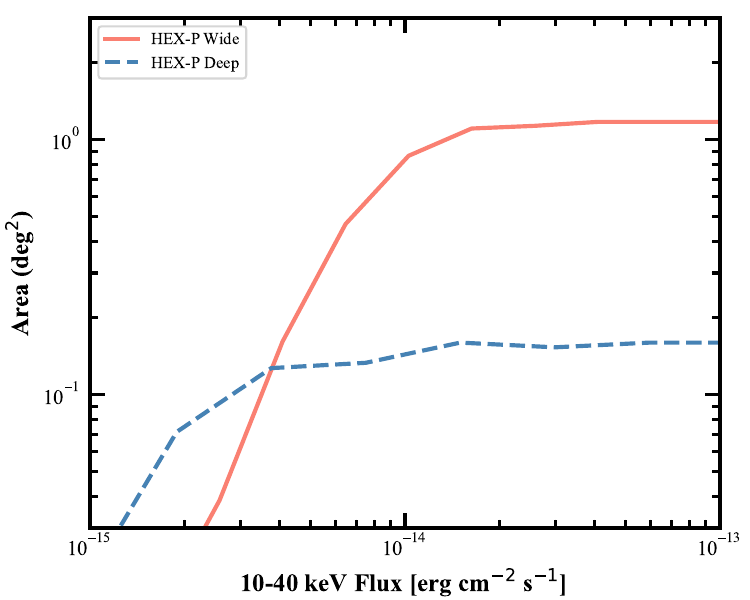}
\includegraphics[width=0.45\textwidth]{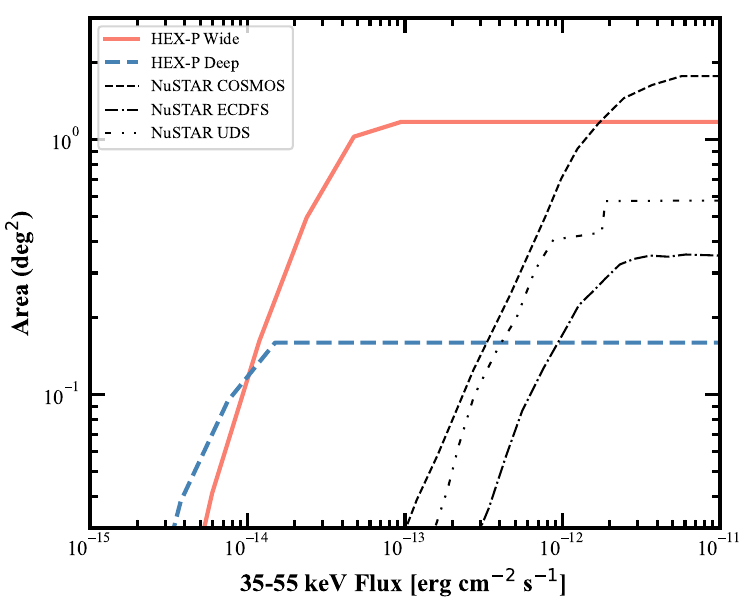}
\caption{\label{sensitivity35}Survey sensitivity for the HET (left) in the 10--40 keV (left) and 35--55 keV (right) bands. The deep and wide surveys are in blue and salmon, respectively. The dashed, dot-dashed and dot-dot-dashed lines in the right panel shows the sensitivity in the same band computed for the COSMOS, ECDFS and UDS \nustar\ surveys \citep{masini18b}. }
\end{figure}

Because one of \textit{HEX-P}'s major goals is to detect the sources that contribute to the peak of the CXB, we have also performed the detection in higher energy bands, 10--40 and 35--55 keV, which have never been effectively exploited and explored before. The sensitivities in the two bands are presented in Figure \ref{sensitivity35}. While a previous attempt to detect sources in the 35--55 keV band was performed by \citet{masini18b} combining the COSMOS, ECDFS and UDS \nustar\ surveys, no sources were detected above the reliability threshold due to the very shallow flux limit in that band (brighter than 10$^{-13}$ erg s$^{-1}$ cm$^{-2}$; see Figure \ref{sensitivity35}, right panel). Their derived upper limit on the expected number counts is consistent with population synthesis model predictions (e.g., \citealt{ananna19}, \citealt{gilli07}), \textit{HEX-P} instead will provide detected number counts for the first time up to at least 55 keV to stringently test the model predictions.

\section{Source statistics and properties}
In Table \ref{Table:number_of_sources}, the number of detected sources with DET\underline{\;\;}ML above the thresholds are reported for the wide and deep surveys in five energy bands for the HET and two energy bands for the LET. The total number of sources detected in the HET is $>$20 times the number of sources detected in previous \nustar\ surveys (with similar exposure and area covered), allowing for the first time not only to resolve the sources contributing to the peak of the CXB but also to perform a statistically significant analysis of their X-ray properties and eventually correlate these with the properties of their host galaxies. The advantage of having a high-- and a low--energy telescope is that the majority of the sources detected in the HET will have a lower-energy counterpart, as shown from the numbers in Table \ref{Table:number_of_sources}. This allows for broad-band analysis as well as use the sharper lower energy PSF for multiwavelength associations and characterization.

As mentioned above, the detection in the hardest X-ray band (35--55\,keV) did not find any sources in previous \nustar\ surveys. In the \textit{HEX-P} surveys, we detect more than $\sim$1000 sources in the 10--40 keV band and $\sim$70 sources in the 35--55 keV band, combining deep and wide surveys. These results show that \textit{HEX-P} will be able to provide the first detections in a band encompassing the entire peak of the CXB and above it.

\begin{figure}
\centering
\includegraphics[width=0.49\textwidth]{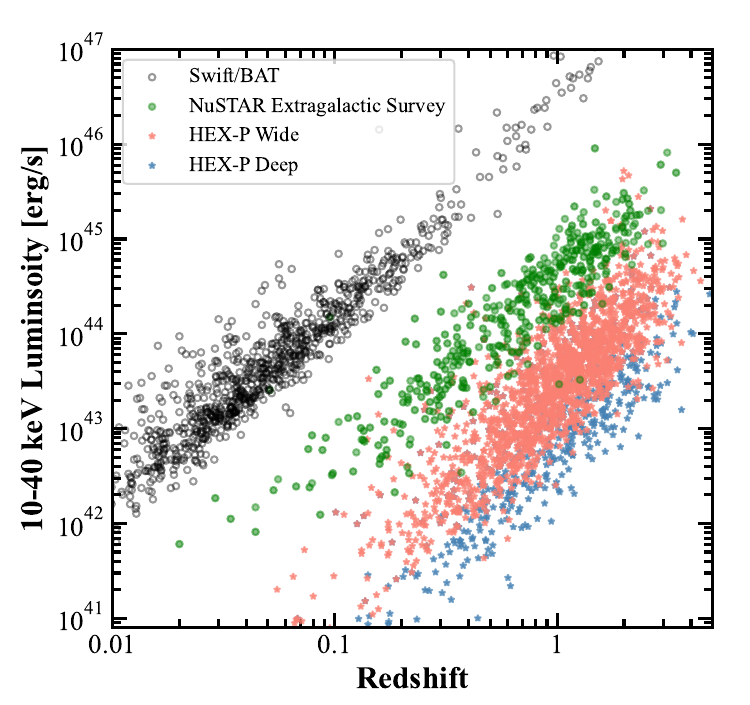}
\includegraphics[width=0.49\textwidth]{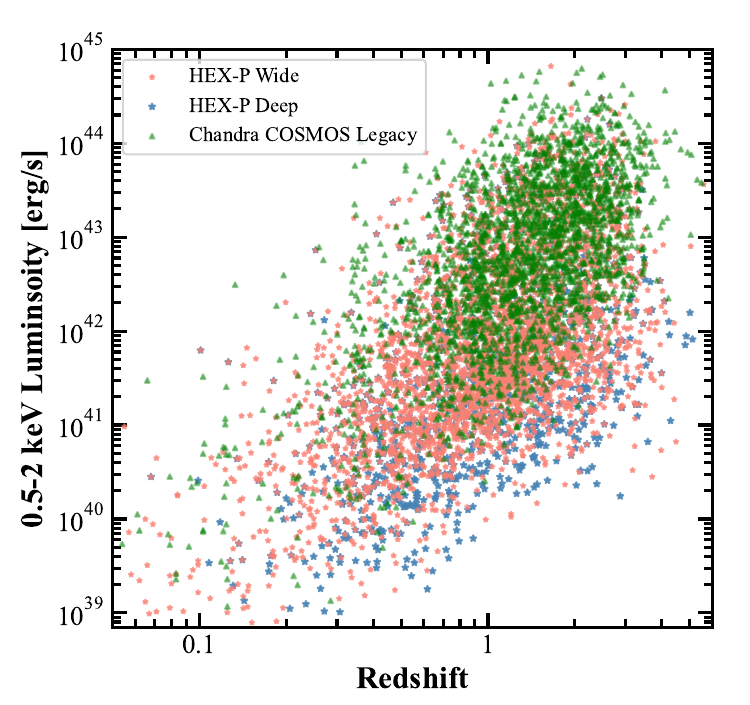}

\caption{\label{lxz} Left: 10--40\,keV rest frame luminosity versus redshift of the HET detected sources in the deep (blue) and wide (salmon) surveys. Right: 0.5--2\,keV rest frame luminosity versus redshift of the LET detected sources in the deep (blue) and wide (salmon) surveys. In both panels, the \textit{HEX-P} surveys are compared with surveys in the literature using \nustar\ and \textit{Swift}-BAT in the left panel and \chandra\ in the right panel.}
\end{figure}

Thanks to the information in the original mock catalog used for populating the sources of the \textit{HEX-P} simulations, we associate a redshift and a column density to each detected source. By the time \textit{HEX-P} is launched, extensive spectroscopic campaigns and photometric survey data will be available in the majority of the well-known fields, using data from {\it JWST}, {\it Euclid}, {\it Roman} as well as ground-based telescopes like Rubin, 4MOST and SDSS. This will allow the complete characterization of the multiwavelength properties of \textit{HEX-P}--detected AGN. In Figure \ref{lxz}, the HET and LET sources from the deep and wide surveys are plotted. The reported luminosities were not corrected for intrinsic absorption. The \textit{HEX-P} detections are compared with samples in the literature including the \textit{Swift}-BAT 105 month all-sky survey sample \citep[black open circles;][]{Oh_2018}, the catalog of \nustar\ extragalactic survey sources, including COSMOS \citep{civano15}, ECDFS \citep{mullaney15}, UDS \citep{masini18}, and the 40-month serendipitous \citep{Lansbury_2017} surveys.
While the \nustar\ surveys already reach luminosities two orders of magnitude fainter than the \textit{Swift}-BAT sample and extend to significantly higher redshift, with \textit{HEX-P} sources, it will be possible to explore a new area of the high-energy luminosity-redshift space by reaching Seyfert-like luminosities up to redshift $z\sim3$ and for the first time to have a statistically significant sample of sources at $z>2$, thanks to the combination of flux and area covered. Even in the low energy band, the LET surveys will achieve luminosity and redshift limits comparable or deeper than archival \chandra\ and \xmm\ surveys.

\section{CXB resolved fraction}

Before \nustar\, the spectrum of the CXB in the hard X-ray band ($>$10 keV) was used as an integral constraint due to the non-focusing nature of hard X-ray telescopes. 
As mentioned in Section 1, thanks to \nustar\ we are able to resolve, for the first time, $\sim$30--35\% of the CXB above 8 keV. 

With \textit{HEX-P}, it will be possible to measure the resolved fraction up to energies of $\sim$40 keV. 
Because we used the well-calibrated mock catalog from \citet{marchesi20}, computing the resolved fraction in the simulated \textit{HEX-P} surveys is straightforward. 

To this end, we have summed the input fluxes of the HET--detected sources in the deep and wide surveys separately in four different bands (3--8~keV, 8--16~keV, 8--24~keV, and 10--40~keV) that were chosen a posteriori to minimize the uncertainty on the resolved fraction. We then summed the fluxes of all the sources in the total input mock catalog in each energy band, which indeed returns the total intensity of the CXB (within a few percent) as reported and shown in Figure 5 of \citet{marchesi20}. The CXB resolved fraction in the wide and deep surveys was then calculated by dividing the sum of the input fluxes of the detected sources by the total fluxes from the input catalog in each of the four aforementioned energy bands. 

The uncertainty on the resolved fraction was derived from the uncertainty on the measured net counts of the detected sources. The net counts of each detected source are the background subtracted counts, extracted from a circular region centered at the source detected position with a 8.5\arcsec\ radius (corresponding to an energy encircled fraction of 50\%). The uncertainty on the net counts ($\sigma_{\rm net}$) of each source is the combination of the uncertainties on the total ($\sigma_{\rm tot}$) and background counts ($\sigma_{\rm bkg}$) in quadrature ($\sigma_{\rm net}^2$ = $\sigma_{\rm tot}^2$ + $\sigma_{\rm bkg}^2$). The uncertainties on the total and background counts are calculated using equation (9) in \citet{Gehrels1986}, which is optimized for sources with small numbers of observed photon counts. Here, we use the 90\% confidence level uncertainties. The net count uncertainty of the entire sample of detected sources ($\sigma_{\rm net,sample}^2$) is calculated by adding up the net count uncertainty of each detected source in the sample in quadrature ($\sigma_{\rm net,sample}^2$ = $\sum\sigma_{\rm net}^2$). The uncertainty on the CXB resolved fraction divided by the CXB resolved fraction is thus the net count (or flux) uncertainty of the entire sample divided by the net count (or flux) of the entire sample.

\begin{figure}
\centering
\includegraphics[width=0.49\textwidth]{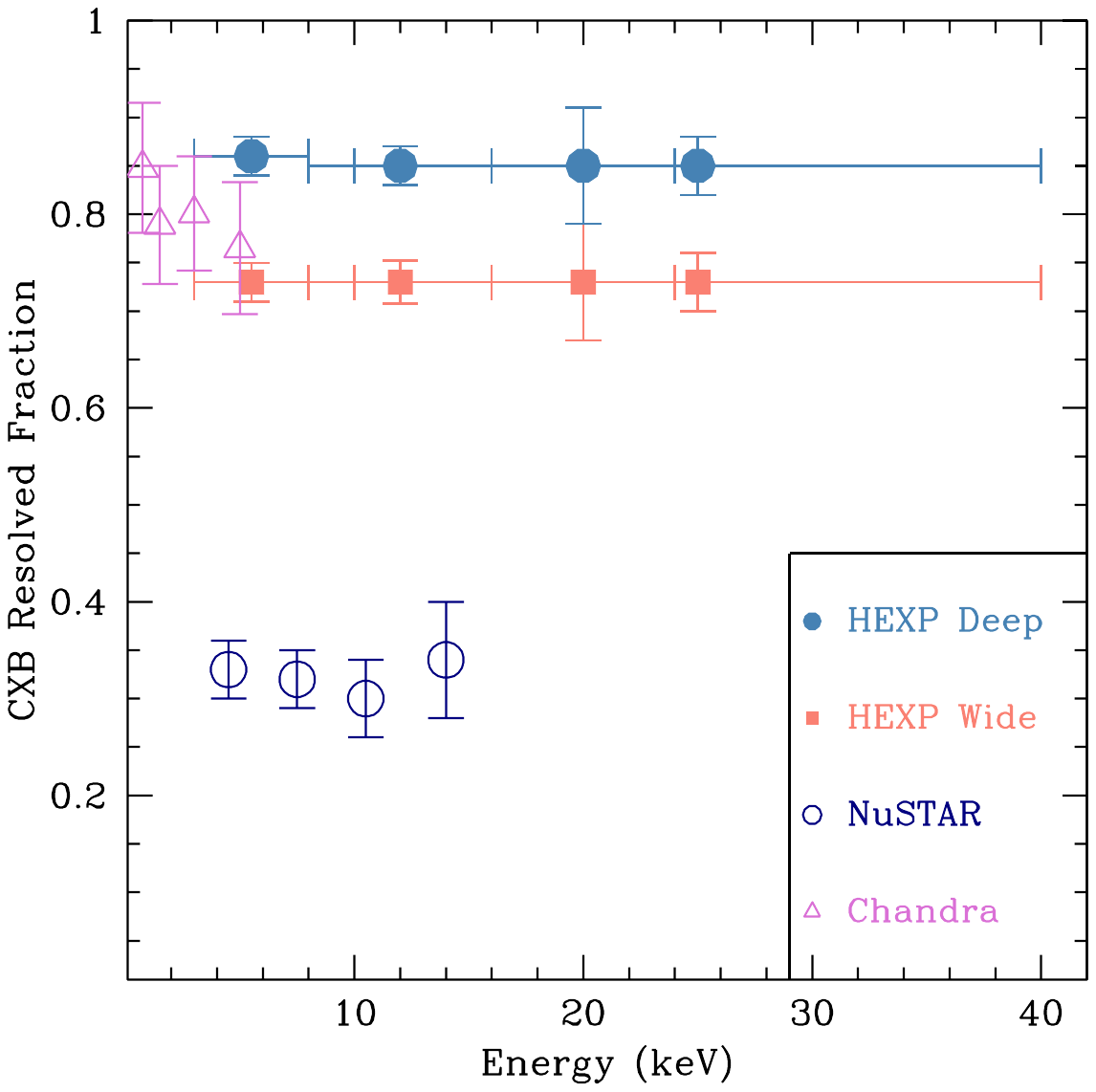}
\includegraphics[width=0.49\textwidth]{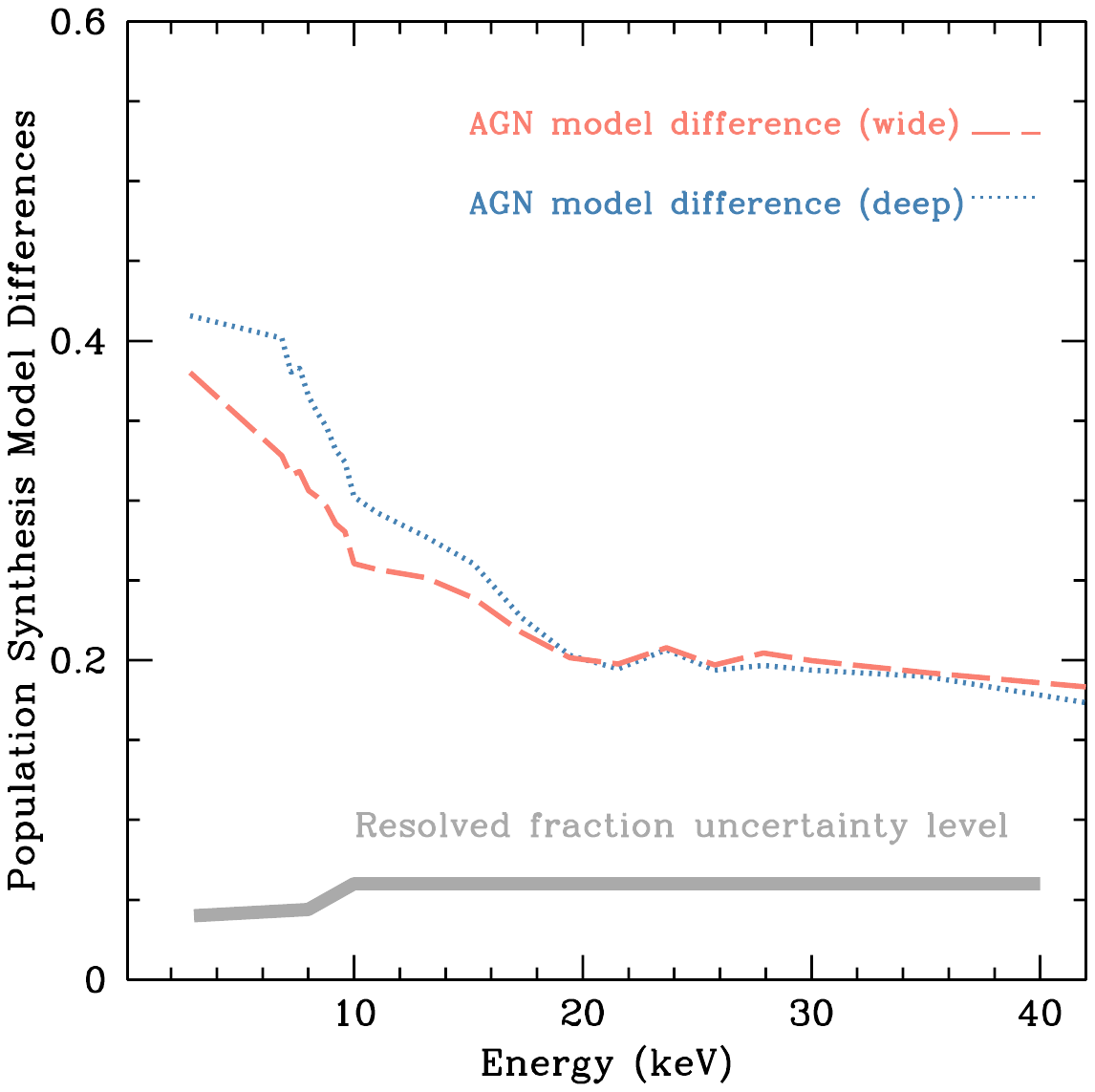}
\vspace{-2cm}
\caption{\label{cxb_fract} Left: Resolved fraction of the CXB as a function of energy as measured by \chandra\ (open violet triangles), \nustar\ deep surveys (open blue circles) and predicted for \textit{HEX-P} deep (solid teal circles) and wide (solid salmon squares). Right: Difference between population synthesis models compared to the level of uncertainties (grey) measured for the CXB resolved fraction. The population synthesis models from \citet{gilli07} and \citet{ananna19} are folded with the deep (teal) and wide (salmon) survey sensitivity curves.}
\end{figure}

As a result, the measured CXB resolved fractions (and errors) in the wide survey are 73\%$\pm$2\%, 73\%$\pm$2\%, 73\%$\pm$6\%, and 74\%$\pm$3\% in the 3--8, 8--16, 16--24, and 10--40~keV energy bands, respectively and the measured CXB resolved fractions (and errors) in the deep survey are 86\%$\pm$2\%, 85\%$\pm$2\%, 86\%$\pm$6\%, and 86\%$\pm$3\% in the four energy bands, respectively. The measured CXB resolved fraction is shown in Figure \ref{cxb_fract} together with the \chandra\ and \nustar\ measurements. The width of the x-axis error bar on each of the \textit{HEX-P} point corresponds to the width of the energy band used for that measurement. 

Just using the detected sources in the HET, \textit{HEX-P} is able to reach a resolved fraction comparable with the results obtained from \chandra\ deep surveys below 8 keV \citep[e.g.,][]{xue12stack} but reaching above the peak of the CXB up to 40 keV (see Figure \ref{cxb_fract}). While $\sim$70 sources are detected in the 35--55 keV band as shown in Table \ref{Table:number_of_sources}, this sample is too small to obtain strong constraints on the resolved fraction and provides only an upper limit (not reported in Figure \ref{cxb_fract}). While the actual intensity of the CXB remains still uncertain, as recently shown by \citet{Rossland23}, the resolved fraction measured here is self-consistent as the input source fluxes are drawn from the mock catalog of \citet{marchesi20} and the total flux used is the sum of all the sources from the same mock catalog. Therefore, the uncertainties affecting the intensity of the CXB do not play a role in our measurement.

A fraction of the detected sources might be affected by Eddington bias, and their measured fluxes might be over-estimated compared to the input fluxes. This excess (over-estimation) of the measured fluxes mostly affects the sources at the flux limit of a survey \citep[see, e.g., Figure 11 in][]{zhao21}. In this paper, the input fluxes from the mock catalog were used to compute the resolved fraction, and consequently, the Eddington bias does not affect our measurement. However, we did assess how Eddington bias will affect the measured resolved CXB fraction in real \textit{HEX-P} surveys compared to the resolved CXB fraction computed above. For this analysis, we used the 50\%-area sensitivity reported in Table 1 as the indicator of the turnover in a flux-flux distribution, following the method in \citet{zhao21}, and derived that fluxes below the 50\% area sensitivity in the wide survey might be overestimated by 2\%, 6\%, 10\%, and 3\% for the 3--8, 8--16, 16--24, and 10--40 keV bands. On the other hand, the fluxes of the sources in the deep survey are predicted to be overestimated by less than 1--2\%.  Therefore, we anticipate that the resolved CXB fraction computed from future \textit{HEX-P} surveys might be higher than the resolved CXB fraction presented in this paper by 2--10\% in the wide survey (depending on the energy band) and $<$2\% in the deep survey.

\subsection{Comparison with population synthesis models}



As shown in Figure~\ref{lxz}, the luminosity-redshift space sampled by \textit{HEX-P} will fill an important gap in our understanding of the AGN population. 
Currently, there are significant disagreements between population synthesis models, in particular at the faintest fluxes where the lack of detected sources makes it impossible to constrain the models. The faintest fluxes are populated by low-luminosity and/or heavily obscured AGN (see Figures~12 and 15 from \citealp{ananna19}). Sampling these parts of the population is one of the goals of the \textit{HEX-P} surveys and will be of extreme importance for calibrating the next generation of synthesis models. 

The wide survey will also provide high-energy information on the poorly sampled population of luminous AGN which are rare and require larger volumes to be found (see the right panel of Figure~\ref{lxz}). Although different population synthesis models predict different fractions of AGN at each obscuration level, usually these contributions sum to the same value of CXB (e.g., the CXB broken into bins of obscuration is given in Figure 11 of \citealp{ananna19}). One way to break the degeneracy is by applying different flux limits in different energy bands, as that leads to varying predictions between models of the CXB resolved fraction. 

Comparing the measured resolved fraction with population synthesis models in this work would not return valuable information as the mock catalog used in these simulations is drawn from the \citet{gilli07} model and therefore the obtained results are in agreement with it. However, we can use the measurement uncertainties to showcase \textit{HEX-P}'s capability to distinguish between model differences. In Figure \ref{cxb_fract} (right panel), the uncertainty level measured on the CXB resolved fraction (gray band) is compared with the typical difference between resolved fractions predicted by population synthesis models at the \textit{HEX-P} flux limits in the wide and deep surveys. We used the models from \citet{gilli07} and \citet{ananna19} for this comparison, all of which were folded with the deep and wide survey sensitivities. Several other models were considered, but the above models captured the typical discrepancy between model predictions. We find that the difference between models is significantly larger than the simulated \textit{HEX-P} measurement uncertainties, which will help us converge toward the most accurate population synthesis model.


 \textit{HEX-P} will focus X-rays in an energy range where it has never been achieved by any instruments before, allowing us to study the hard X-ray emission of AGN to flux limits never reached before. This will open the parameter space for new serendipitous discoveries. It is possible, even likely, that no existing model will perfectly predict the population that \textit{HEX-P} will observe, as has happened in the past when new facilities reached unprecedented capabilities \citep{ananna19,kirkpatrick2023}. 


\section{Obscured sources in surveys}


Besides the detection of the sources contributing to the peak of the CXB, one of the main goals of \textit{HEX-P}
extragalactic surveys will be to characterize AGN spectral properties such as spectral
indices, cut-off temperatures, reflection scaling factors and circum-nuclear obscuration covering factors \citep{Kammoun23,Piotrowska2023,boorman23}.
Extragalactic field surveys will contribute to this analysis by providing deep exposures for faint sources. While the CXB is an aggregate population statistic, data from \textit{HEX-P} can be directly incorporated into models and can contribute greatly towards constraining one of the biggest uncertainties in AGN populations: the fraction of heavily obscured (i.e., CT) objects.


\cite{zap18} carried out a systematic broad-band (0.5--24 keV) spectral analysis of 63 sources detected in \nustar\ extragalactic surveys with a flux brighter than 7$\times$10$^{-14}$ \cgs\ (8--24 keV) to characterize their spectral properties, obtain a column density distribution and measure the absorbed/CT fractions to compare with predictions from population-synthesis models. However, their sample lacks large statistics, its soft band data (needed to perform the spectral fitting) was taken at a different epoch and the spectral analysis can be affected by variability. As a consequence, the results on the derived properties, like the obscured and CT fractions, have large uncertainties or are basically unconstrained (CT = 0.02--0.56\, or $<$0.66 at 90\% c.l.). Clearly, increasing the number of sources with good quality spectra in the hard X-ray band paired with simultaneous good quality low-energy data would be crucial 
(as shown in \citealp{marchesi18,marchesi19}) to constrain the CT fraction at a flux limit never reached before. \textit{HEX-P} surveys will provide such a sample. 

\begin{figure}
\centering
\includegraphics[width=0.43\textwidth]{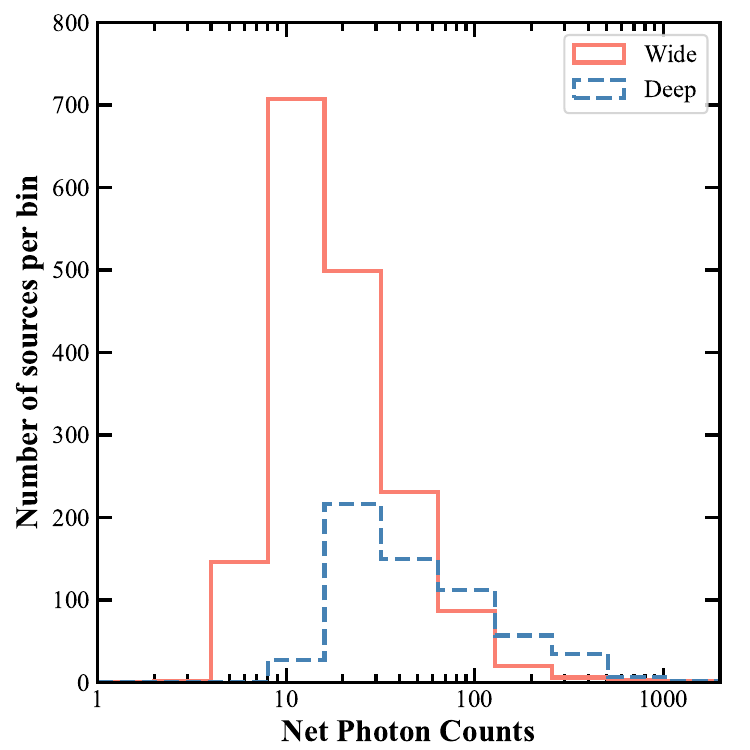}
\includegraphics[width=0.43\textwidth]{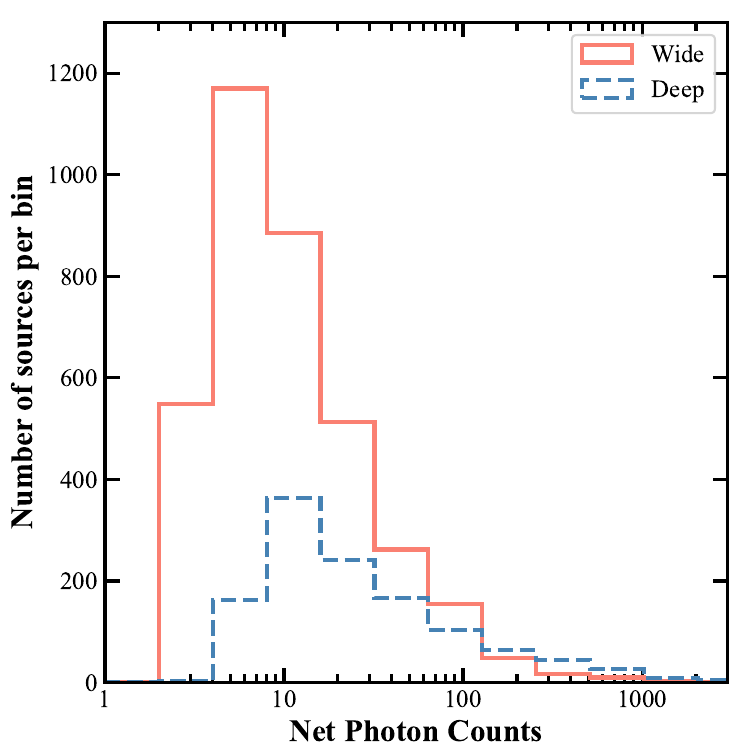}

\caption{\label{counts} Source count distributions showcasing how many counts we will have in HET (left) and LET (right) over the bandpass to perform spectral analysis. The counts were extracted in the 0.5--15 keV band for the LET and the 3--40 keV band for the HET (summing two telescopes). The sources from the deep survey are in blue, those from the wide survey are in salmon.}
\end{figure}

In Figure \ref{counts}, the net counts distribution in 10'' and 5'' radius apertures ($\sim$30\% of the PSF) for the HET and LET detections are reported. As shown by \citet{zap18}, above 40 counts in the 3--24 keV band, it is possible to obtain good ($>$3$\sigma$) constraints on spectral parameters with spectral fitting.  The sample in \textit{HEX-P} surveys will include several hundreds of sources with broadband coverage and enough counts in the 0.5--40 keV energy band to perform good quality spectral analysis (see Section 6.1). Besides the obscuration distribution, for the bright sources in the sample, it will also be possible to measure other spectral properties like the high-energy cut-off as shown by \cite{Kammoun23} for both unobscured and obscured AGN (see their Section 5 and Figures 5 and 6) extending their results to fainter luminosities/fluxes.

The good quality X-ray spectra will also allow us to investigate further potential observational biases caused by rapidly spinning black holes with strong reflection components that mimic obscured spectra \citep{Gandhi07, Vasudevan16}. Deconvolving spin-related reflection signatures from obscuration will reval SMBH growth with cosmic time.

\begin{figure}
\centering
\includegraphics[width=0.49\textwidth]{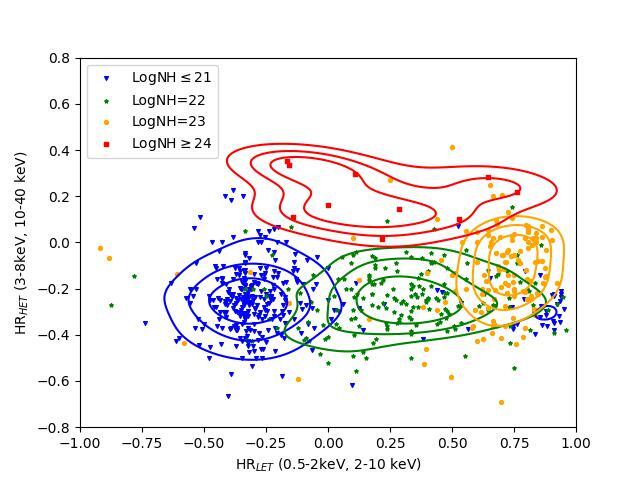}
\includegraphics[width=0.49\textwidth]{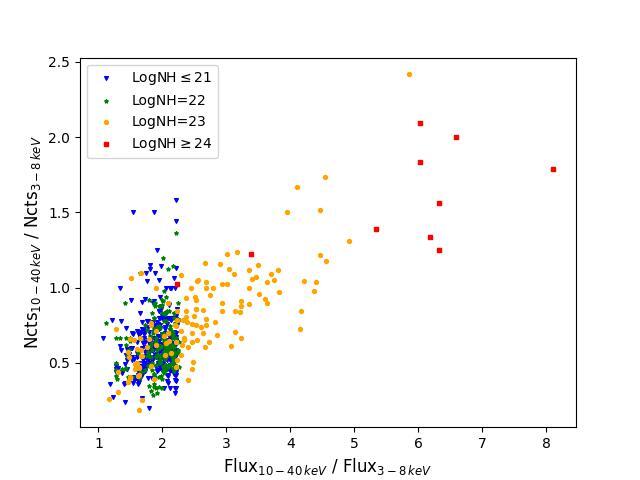}

\caption{\label{hr} {\it Left:} HR$_{HET}$ versus HR$_{LET}$ for sources detected in the 10--40 keV band. The sources are color-coded by obscuration from the input catalog. The density contours represent 1, 2 and 3 $\sigma$ from a two-dimensional Gaussian distribution. {\it Right:} 10--40 and 3--8 keV observed counts ratio plotted against the input flux ratio in the same bands. The sources are color-coded by obscuration from the input catalog.}
\end{figure}

While spectral analysis is the best tool to measure intrinsic
source properties (spectral index, column density, luminosity), the hardness ratio (HR=H-S/H+S where H and S are typically the number of counts in a given hard and soft band respectively) has been used before in \xmm, \chandra\ and \nustar\ surveys \citep{civano15,marchesi16a,zhao21} to obtain a rough measurement of obscuration. As shown in \cite{zhao21}, the hardness ratio computed using \nustar\ bands (3--8 and 8--24 keV) is more sensitive to obscuration above $N_{\rm H} > 10^{24}$ cm$^{-2}$ than just using softer bands (0.5--2 and 2--10 keV), which can better distinguish column densities below $N_{\rm H} < 10^{23}$ cm$^{-2}$. However, usually, soft and hard X-ray data are not simultaneous in extragalactic surveys, and therefore the estimate of obscuration using non-simultaneous hardness ratios can give erroneous results. 
With \textit{HEX-P}, HET detections will have soft energy data readily available to build a color-color diagram spanning the 0.5--40 keV band. Figure \ref{hr} (left panel) shows such a color-color diagram using the hard X-ray HR$_{HET}$ (with H=10--40 keV and S=3--8 keV counts) compared to the soft X-ray HR$_{LET}$ (with H=2--10 keV and S=0.5--2 keV counts). The sources are color-coded according to their column density from the input mock catalog. Only detections in the 10--40 keV and 3--8 keV bands are used. As most of these sources will be in the low-counts regime, a better estimate of the HR will be made using a Bayesian tool such as BEHR \citep{park06}. 
The spread of the points on the x-axis for each given column density is due to the degeneracy of hardness ratio with redshift. Still, we foresee that most, if not all, of the sources in the \textit{HEX-P} extragalactic surveys will have a spectroscopic or photometric redshift measure. 
The contours represent number densities associated with 1, 2 and 3 $\sigma$ from a two-dimensional Gaussian distribution, according to different input obscuration values. Thanks to the coverage at energies above 10 keV, it will be possible to distinguish the CT AGN from the less obscured ones as the CT AGN are separated by the rest of the sources mainly on the HR$_{HET}$ axis/color. Adding multi-wavelength information (e.g., redshifts, infrared fluxes, etc) will constrain the column density to even higher precision and accuracy (e.g., \citealt{Carroll21,carroll23,silver23}) and therefore deliver a robust sample of CT AGN.

Another method to select candidate obscured and CT sources is to use a simple band ratio (H/S) as done in \nustar\ by \cite{mullaney15}. Previously, an anti-correlation was observed between band ratio (H/S) and count rates in the chosen soft band (S), attributed to the decrease of  soft band counts with increasing obscuration \citep{ueda99,mush2000,ceca99,alexander03,tozzi2001}. We have used the 10--40 keV to 3--8 keV input flux ratio and compared it with the detected counts ratio in the same band to test the reliability of the band ratio to find CT sources (Figure \ref{hr}, right panel). The two quantities correlate when considering the 10--40 keV detected sample with fluxes above the 50\%-area flux limit and with detections in both the 10--40 and 3--8 keV bands. When including upper limits on the 3--8 keV flux, the correlation between the two band ratios is not equally strong due to few unobscured sources, which are just detected in the 10--40 keV band, possibly due to Eddington bias.

\begin{table}
    \centering
    \caption{Properties of targets chosen for spectral analysis simulations. \label{tab:obs}}
    \begin{tabular}{l|ccccc}
         \hline
         Object ID
         & $z$ & \textit{NuSTAR} exp & $log(N_{\rm H})$& 2--10~keV flux & 8--24~keV flux \\
         & & (ks) & (cm$^-2$)  & ($10^{-13}$ erg s$^{-1}$ cm$^{-2}$) & ($10^{-13}$ erg s$^{-1}$ cm$^{-2}$) \\
         \hline \hline
        cnuid330 & 0.04 & 52 & 24.13 & 4.75 & 3.54\\ 
        cnuid272 & 0.67 & 106 &  22.62 & 0.65 & 1.03\\ 
        nuid117 & 0.78 & 22 & 24.61 & 0.25 & 0.88\\ 
        \hline
    \end{tabular}
\end{table}

\subsection{Spectral analysis of obscured sources}

\begin{figure}
    \centering
    \includegraphics[width=0.3\textwidth]{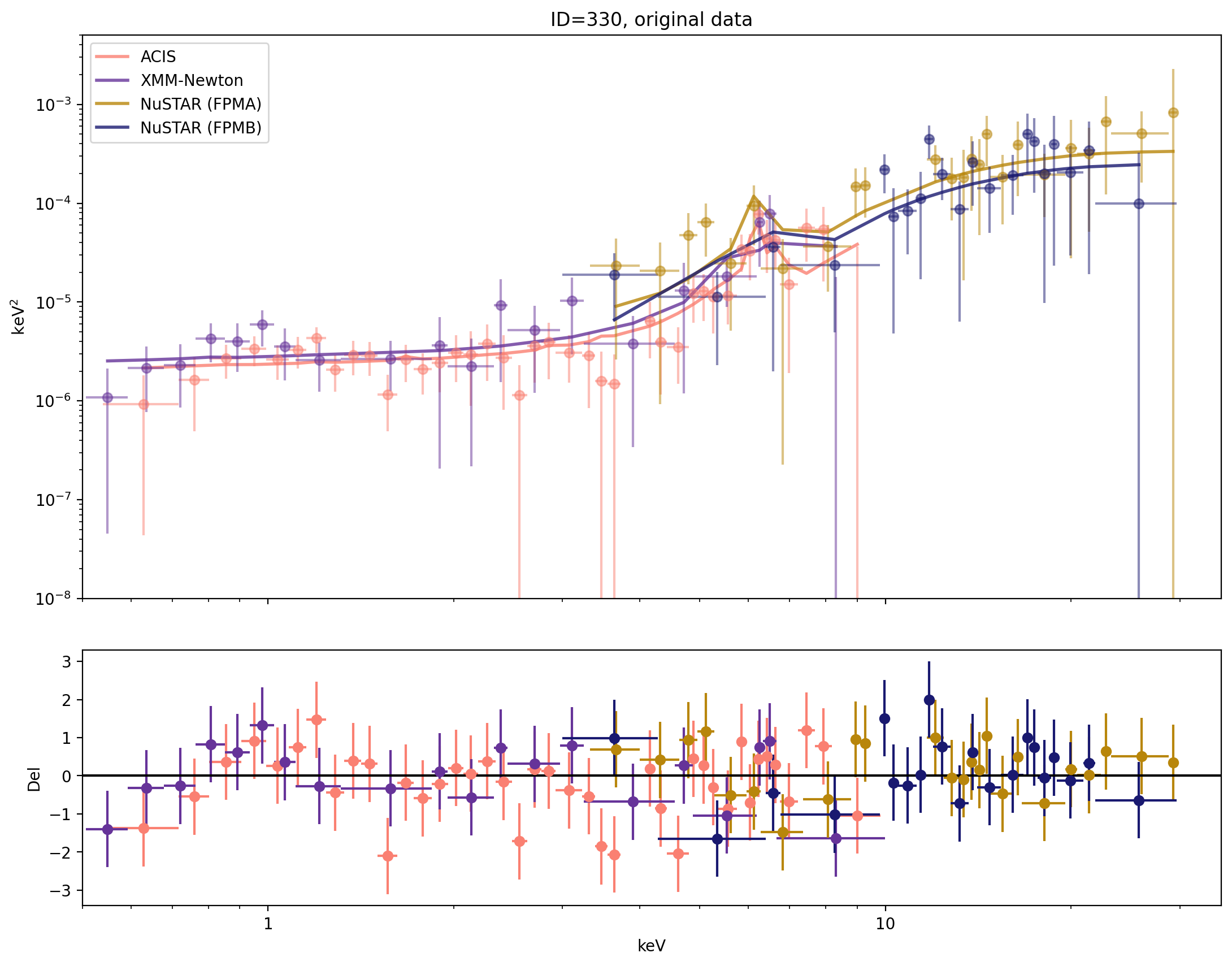}
    \includegraphics[width=0.3\textwidth]{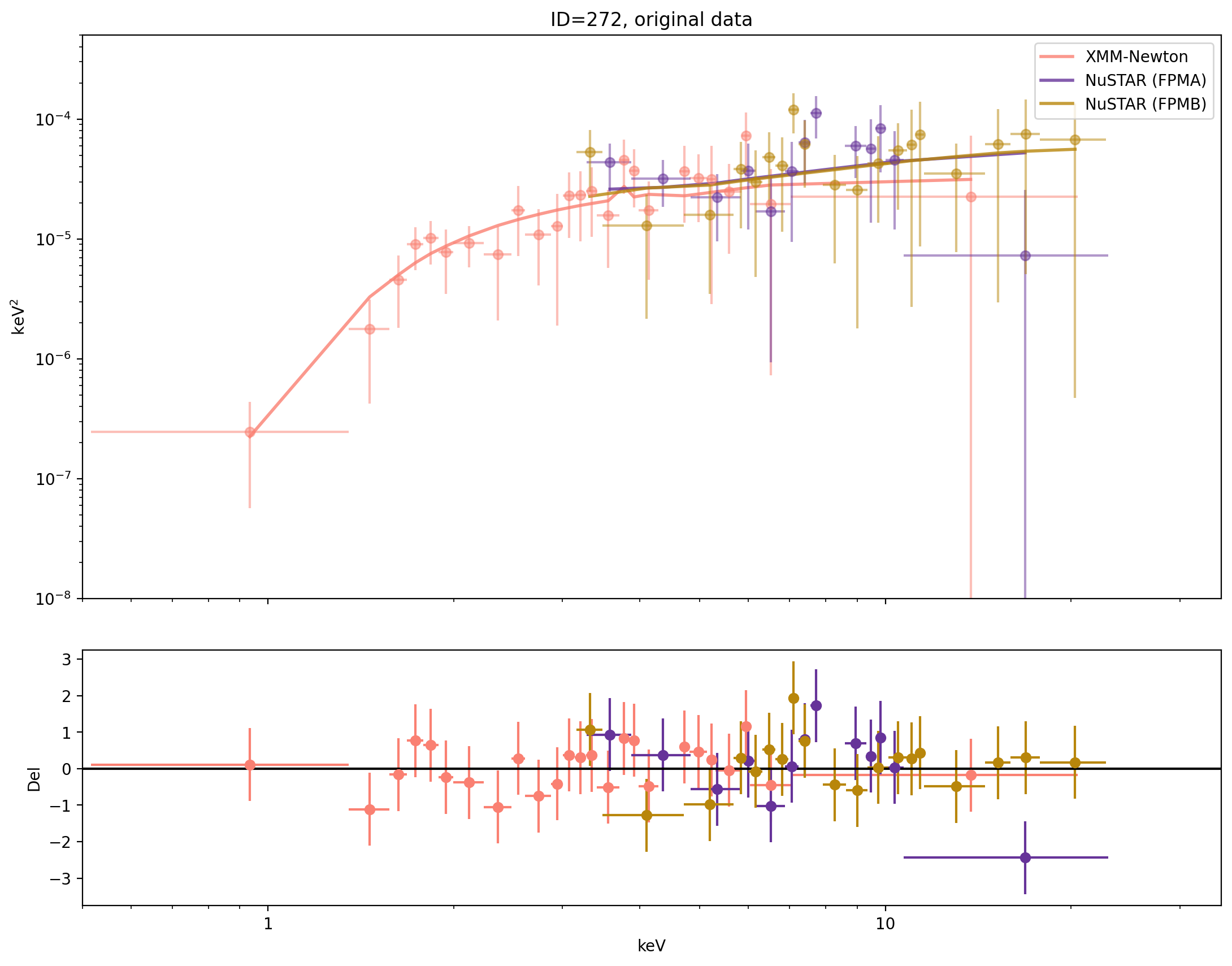}
    \includegraphics[width=0.3\textwidth]{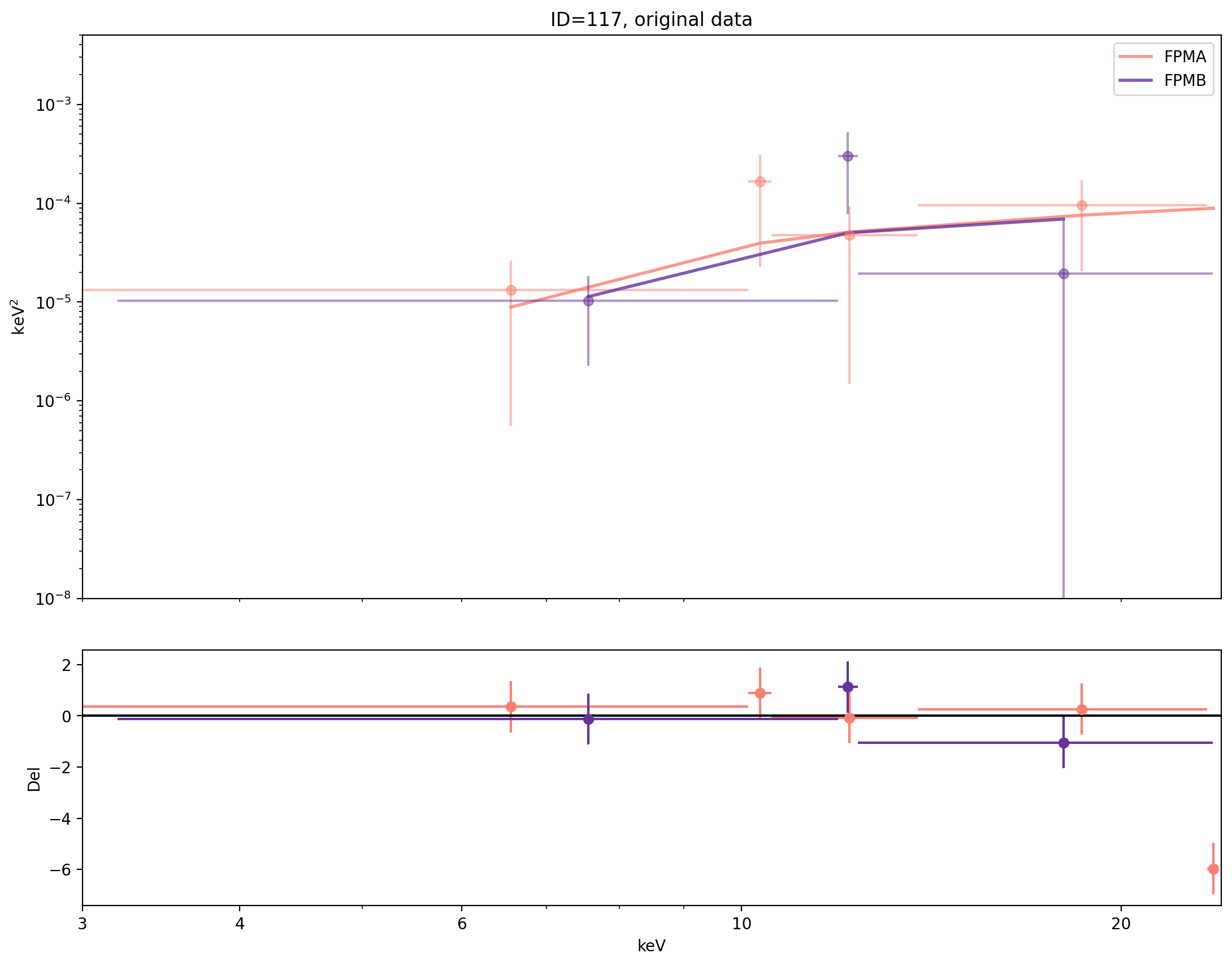}
    \caption{The original spectra for cnuid330 (left), cnuid272 (middle) and nuid117 (right), unfolded with the best-fit \textsc{uxclumpy} model found using BXA denoted with lines.}
    \label{fig:orig_spec}
\end{figure}

To quantify the spectral constraints attainable from \textit{HEX-P} surveys and compare with the quality of the best sources in \nustar's surveys, we simulated known CT and obscured Compton-thin ($log(N_{\rm H})<$24) AGN detected in the \textit{NuSTAR} COSMOS \citep{civano15} and serendipitous \citep{Lansbury_2017} surveys which were both presented in \citet{zap18}. 
The observed/measured spectral properties \citep[from][]{zap18} are summarized in Table \ref{tab:obs}. 
We used the spectra from \cite{zap18} and performed broadband spectral fitting using the Bayesian X-ray Analysis (BXA) code from \cite{BXA} with the \textsc{uxclumpy} model \citep{uxclumpy}. The original \chandra, \xmm\ and \nustar\ spectra and their best-fit models are shown in Figure \ref{fig:orig_spec}. 
As presented by \cite{zap18}, two of the three sources (nuid117 and cnuid330) are classified as CT, and one (cnuid272) is obscured but Compton-thin. 

Using the best-fit \textsc{uxclumpy} models found by BXA, we simulated \textit{HEX-P} spectra for both the LET and HET. For the bright CT source (cnuid330), we simulated exposure times consistent with the proposed wide field survey (50, 60, and 100 ks; see Figure \ref{areaexpo}, salmon curves) which is likely to detect similarly bright CT sources. 
For the two fainter sources (nuid117 and cnuid272), we used exposure times consistent with the \textit{HEX-P} deep survey (250, 500, and 700 ks; see Figure \ref{areaexpo}, blue curves), which probes the faint population. 

\begin{figure}
    \centering
    \includegraphics[width=0.45\textwidth]{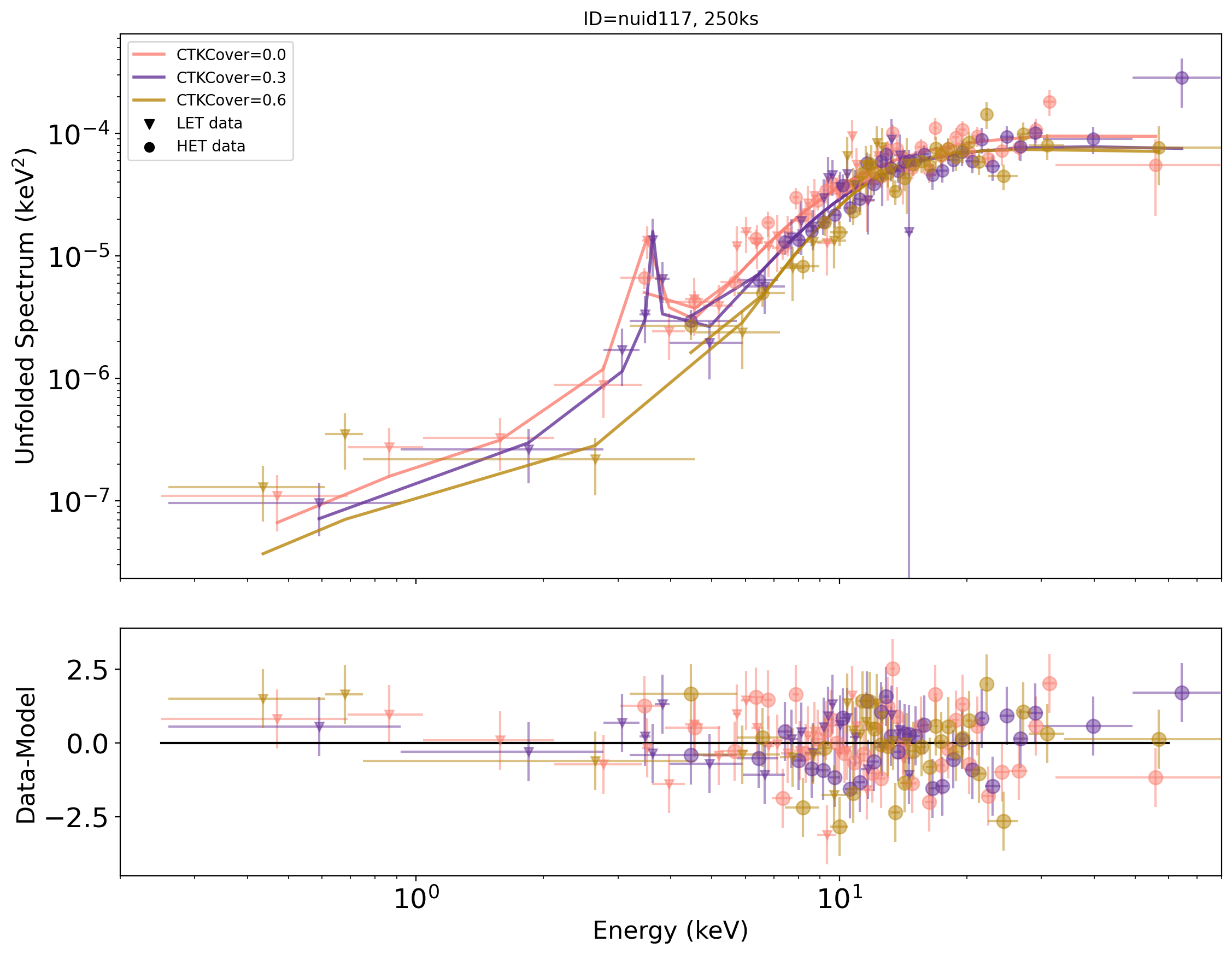}
    \includegraphics[width=0.45\textwidth]{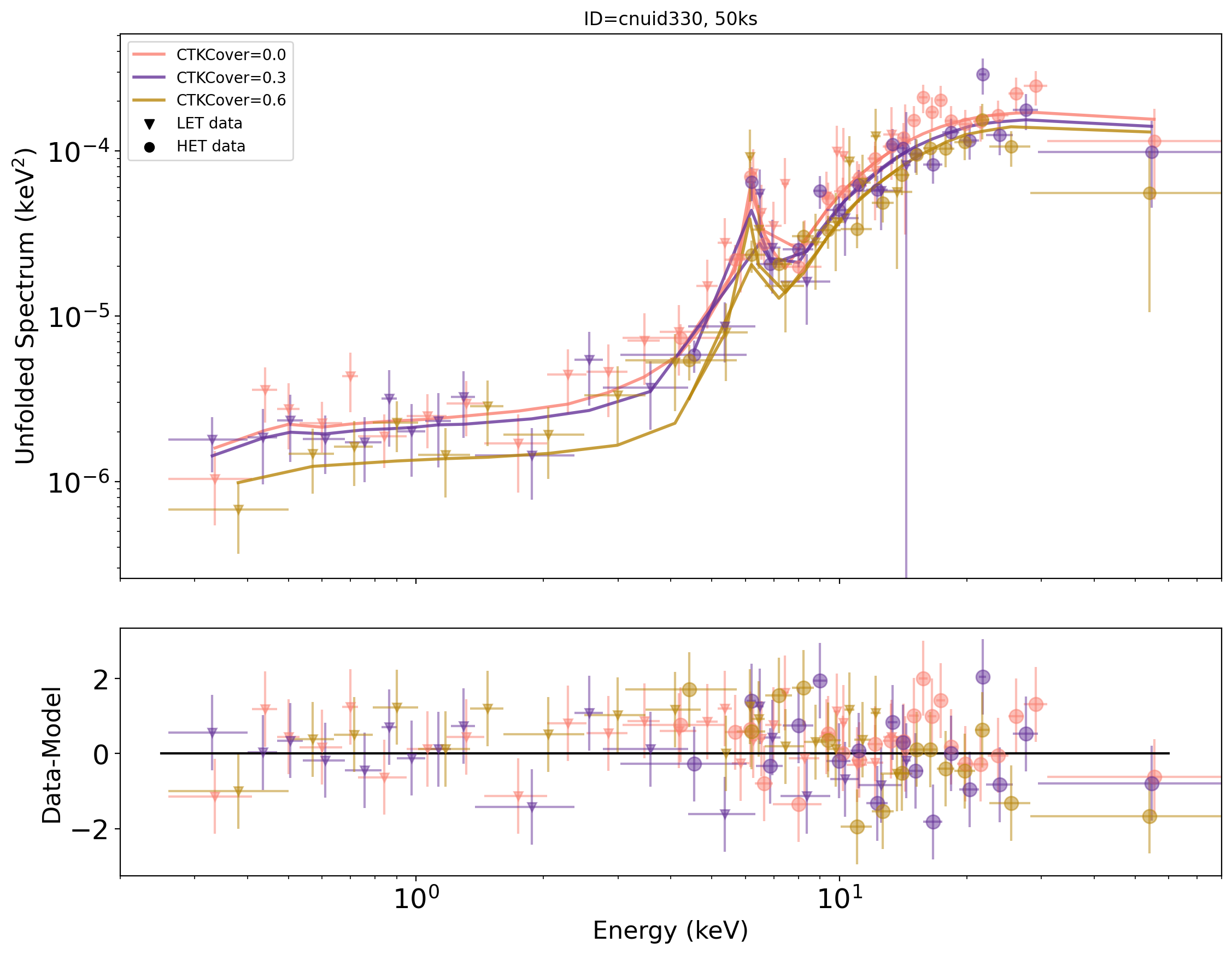}
    \includegraphics[width=0.45\textwidth]{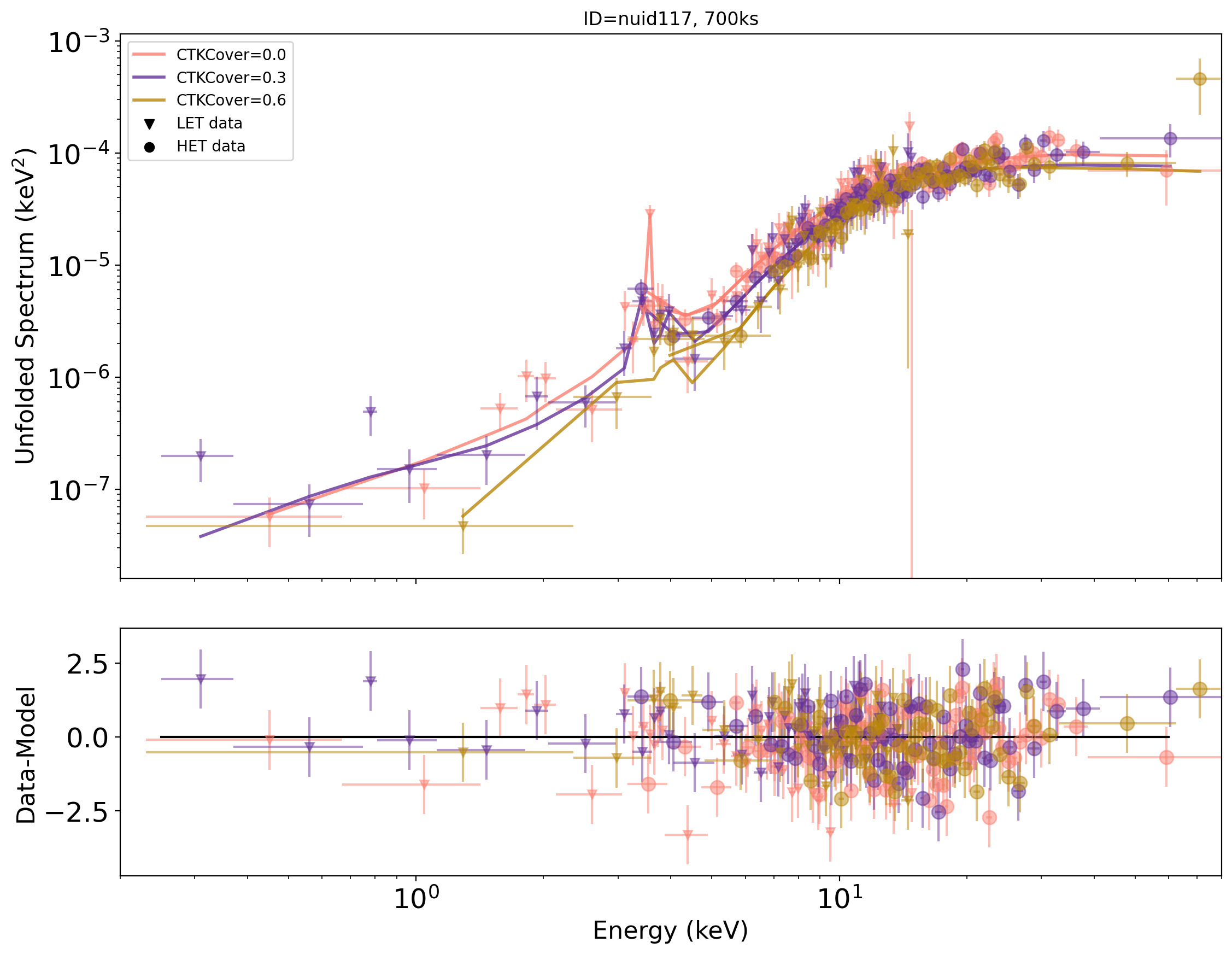}
    \includegraphics[width=0.45\textwidth]{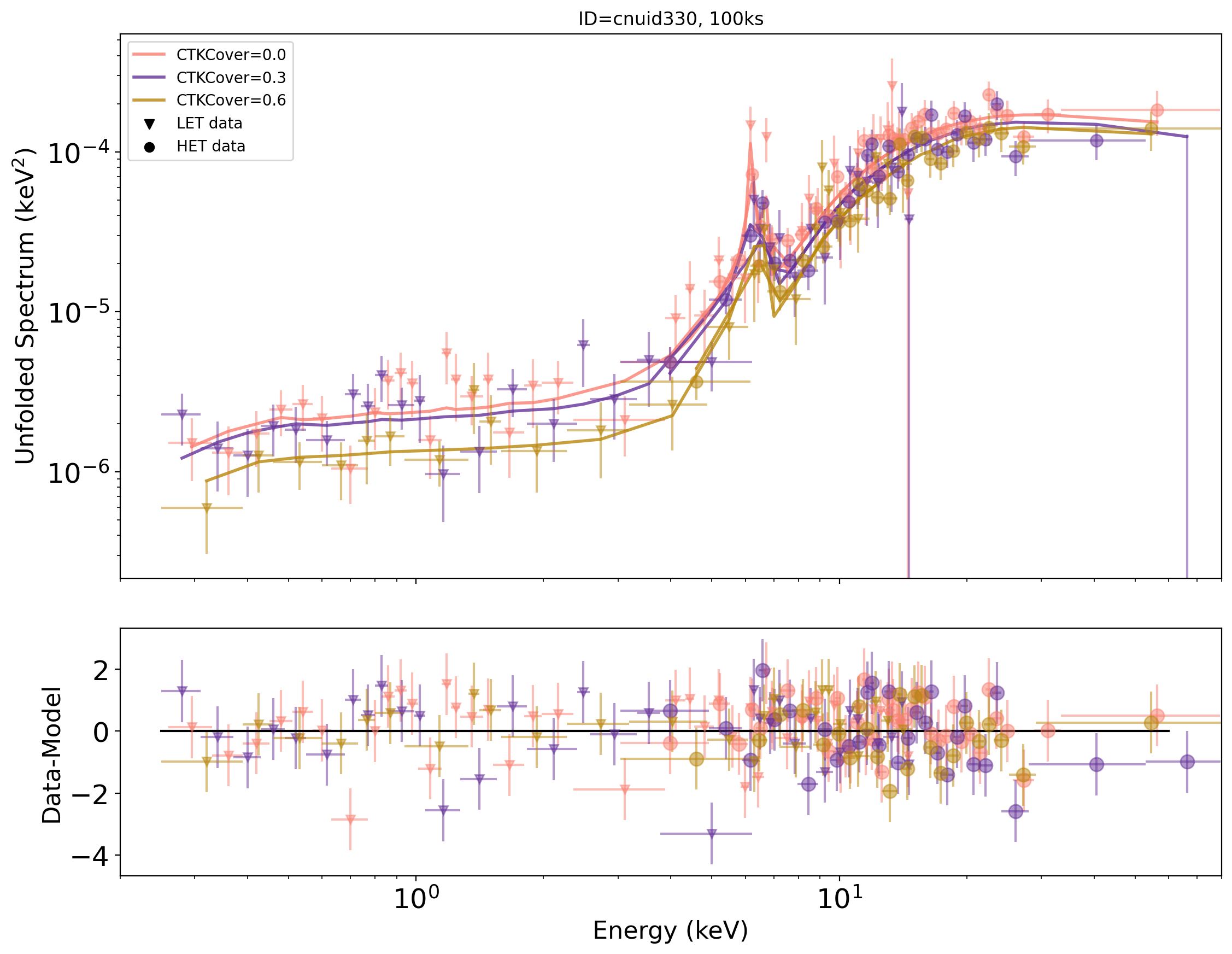}
    \caption{Unfolded \textit{HEX-P} spectra for two CT sources (nuid117, left and cnuid330, right) simulated for the minimum (top) and maximum (bottom) exposure times of the deep (nuid117, left) and wide (cnuid330, right) surveys. LET and HET data are plotted as triangles and circles, respectively. The theoretical spectra on which the simulations were based are shown by solid lines. The three colors denote simulations with different values of \textsc{uxclumpy}'s CTK-cover parameter. nuid117 has been normalized at 7~keV and cnuid330 has been normalized at 5~keV.}
    \label{fig:simspec}
\end{figure}

\begin{figure}
    \centering
    \includegraphics[width=0.49\textwidth]{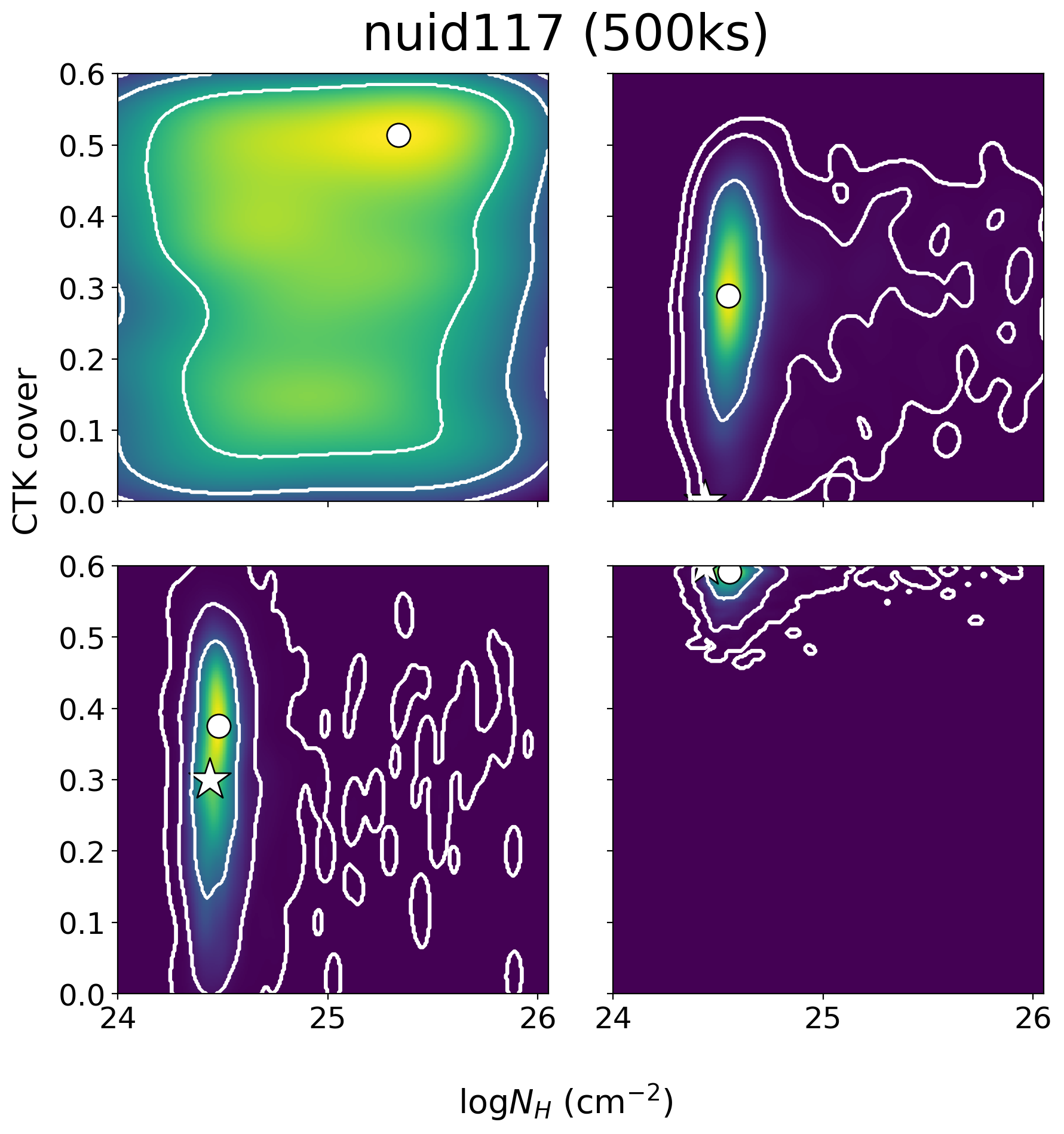}
    \includegraphics[width=0.49\textwidth]{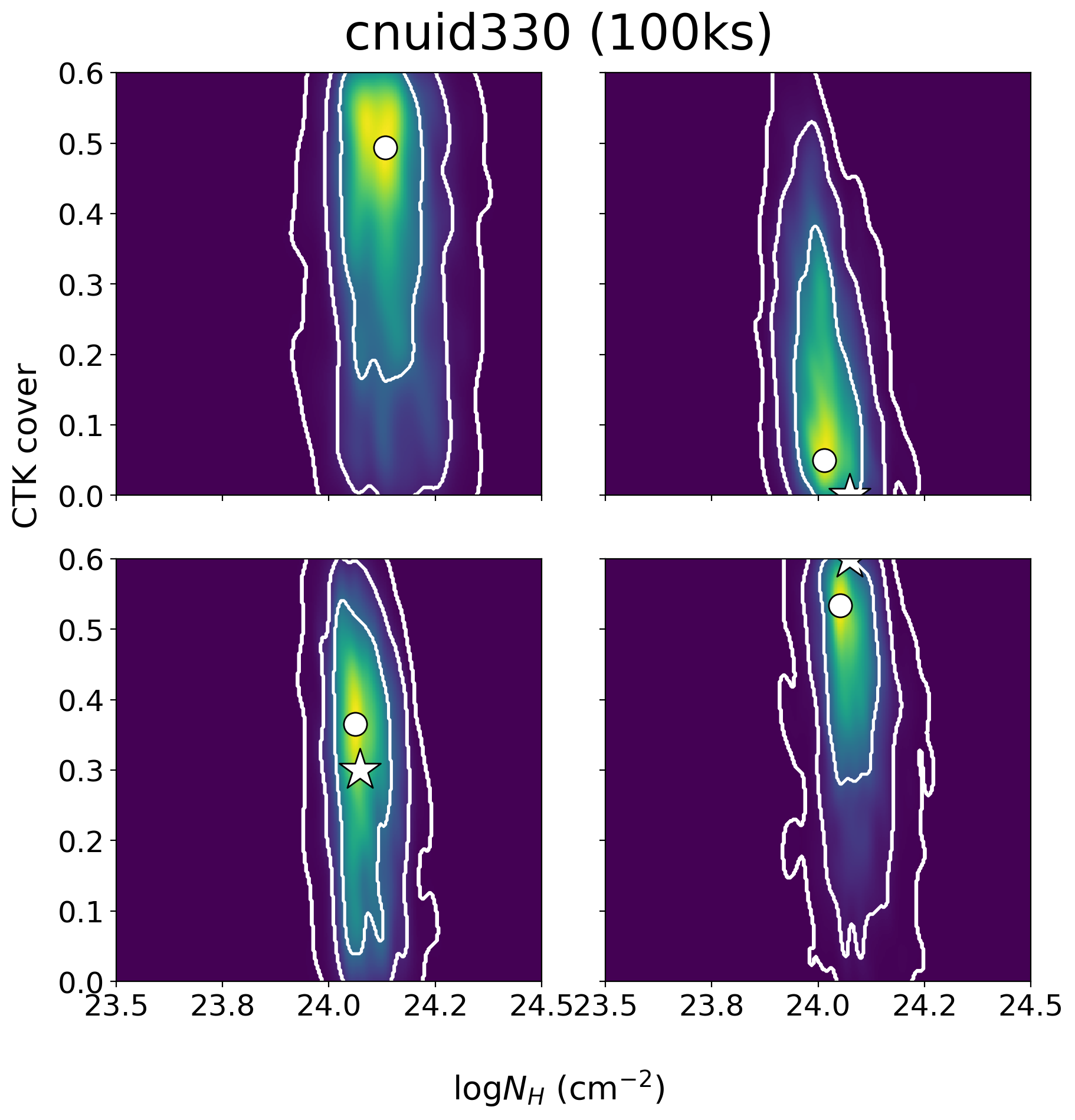}
    \caption{Constraints on $N_{\rm H}$ and \textsc{CTK-cover} for nuid117/cnuid330 (left/right). Constraints are shown for the original data (top-left) and 500 ks/100 ks \textit{HEX-P} simulations with \textsc{CTK-cover}~$=0.0$ (top-right), $0.3$ (bottom-left), and $0.6$ (bottom-right). The white contour lines demonstrate 1$\sigma$, 2$\sigma$, and 3$\sigma$ constraints and a white star denotes the nominal parameter values for the simulations.}
    \label{fig:CT_constraint}
\end{figure}

The goal of these simulations is to test the ability of \textit{HEX-P} to differentiate the subtle effects that the geometry of obscuring material has on the AGN spectral shape.
As described by \citet{uxclumpy}, the \textsc{uxclumpy} model includes an optional CT inner ring of obscuring material that can help to explain the features of some, but not all, CT AGN at hard energies ($\gtrsim 10$~keV). For example, \citet{uxclumpy} shows that this additional component is needed to efficiently model the broadband X-ray spectrum of the prototypical CT-AGN in the Circinus Galaxy.
The inner ring is parameterized by the covering fraction of the material (\textsc{CTK-cover}, which ranges from 0.0 to 0.6). For the faint and higher redshift AGN (nuid117 and cnuid272), \textsc{CTK-cover} is not constrained with previous data. \textit{HEX-P}'s effective area beyond $20$~keV will allow us to discern the subtle effects (e.g., a general sharpening of the Compton hump at $>$10 keV) that \textsc{CTK-cover} has on the hard-band spectra of these AGN.
To show this, we have simulated obscured \textit{HEX-P} spectra with \textsc{CTK-cover}~$= 0.0, 0.3,$ and $0.6$.
The other parameters were set to the best-fit values found by BXA on the original \nustar\ spectra as in Table \ref{tab:obs}. The simulated spectra and models of these objects are shown in Figure \ref{fig:simspec}. 
The effects of different values of \textsc{CTK-cover} for these X-ray faint obscured AGN (nuid117 and cnuid330) are clearly observable in \textit{HEX-P} spectra. The corresponding constraints on \textsc{CTK-cover} are plotted against $N_{\rm H}$ in Figure \ref{fig:CT_constraint} for simulations of nuid117 and cnuid330 using the 500 ks and 100 ks exposures. 
Constraints on \textsc{CTK-cover} and 99.73\% percentile values derived from the posterior distribution for cnuid272 in the deep survey (500~ks) are $0.31 ^{+ 0.19 }_{- 0.30 }$ (true \textsc{CTK-cover}$=0.0$),  $0.37 ^{+ 0.19 }_{- 0.36 }$ (true \textsc{CTK-cover}$=0.3$), and $>0.11$ (true \textsc{CTK-cover}$=0.6$). For cnuid330 in the wide survey (100~ks), the constraints are $0.06 ^{+ 0.496 }_{- 0.064 }$ (true \textsc{CTK-cover}$=0.0$),  $0.37 ^{+ 0.23 }_{- 0.36 }$ (true \textsc{CTK-cover}$=0.3$), and $0.59 ^{+ 0.01 }_{- 0.57 }$ (true \textsc{CTK-cover}$=0.6$). As a side note, it's worth remembering that the spectral analysis discussed here was not conducted in \citet{zap18} as the quality of the spectra did not allow it, while \textit{HEX-P} will provide excellent constraints even on the geometrical properties of the circum-nuclear obscurer itself. spectral covering fraction.

\begin{figure}[b]
    \centering
    \includegraphics[width=0.9\textwidth]{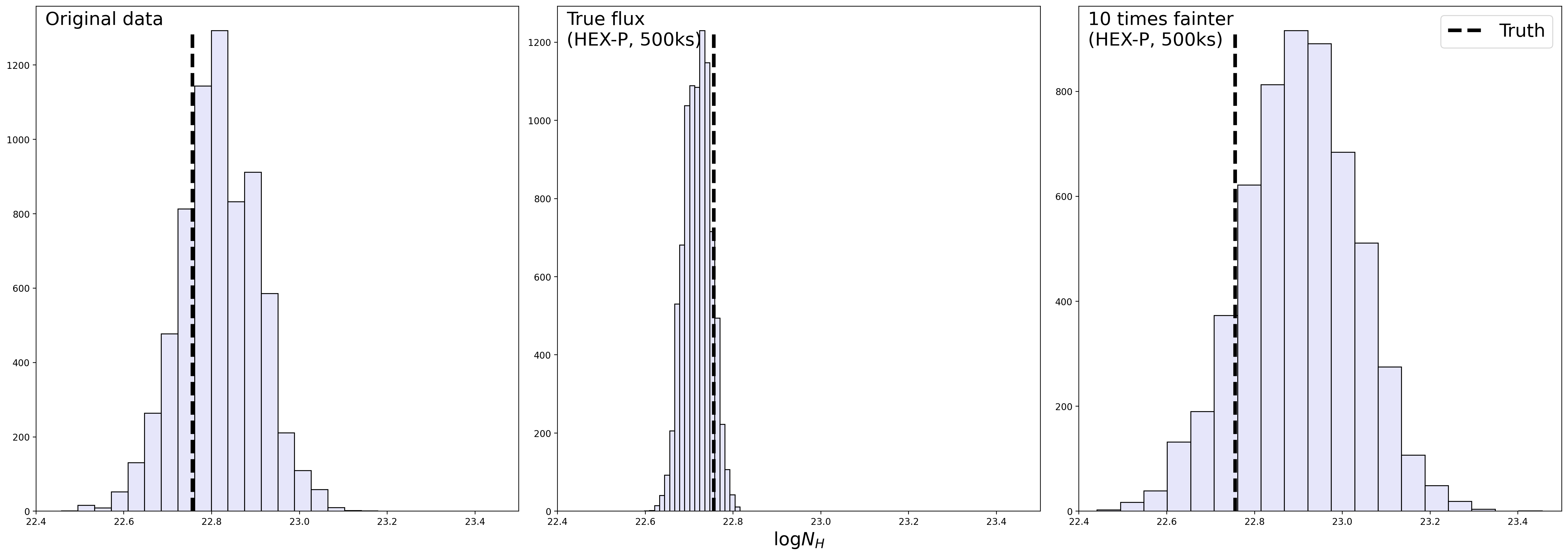}
    \caption{The $N_{\rm H}$ constraints for cnuid272 (a Compton-thin source) found using the original data (left) and 500~ks \textit{HEX-P} simulation (middle) and a simulated object ten times fainter than (but otherwise identical to) cnuid272 (right). The dashed black line in all three panels marks the best-fit $N_{\rm H}$ for the original data, which is set to the ``truth" value for the simulated spectra. }
    \label{fig:272_constraint}
\end{figure}

For obscured, Compton-thin AGN, the effects of \textsc{CTK-cover} are negligible within \textit{HEX-P}'s energy range.
But it is still of interest to constrain $N_{\rm H}$ of these AGN. 
In the case of cnuid272, this constraint is shown for 500~ks simulation of the object at its nominal flux, as well as an identical object that is ten times fainter as well as to the $N_{\rm H}$ constraints of the existing \textit{NuSTAR} observation in Figure \ref{fig:272_constraint}. Similar constraints as \nustar\ will be achieved for a source that is 10 times fainter with \textit{HEX-P}. 
It may be noted that the maximum a posteriori value of $N_{\rm H}$ derived by BXA (which is taken as the ``truth" value for the simulations) does not align with the peak of the distribution. This shift can occur when analyzing a multi-dimensional parameter space where one or more parameters become limits. In these cases, it is difficult to define a best-fit model, and therefore, it is more useful to compare the posterior distribution of the parameters, which is done here.


\subsection{Predictions on the Compton-thick fraction}

\begin{figure}
\centering
\includegraphics[width=0.9\textwidth]{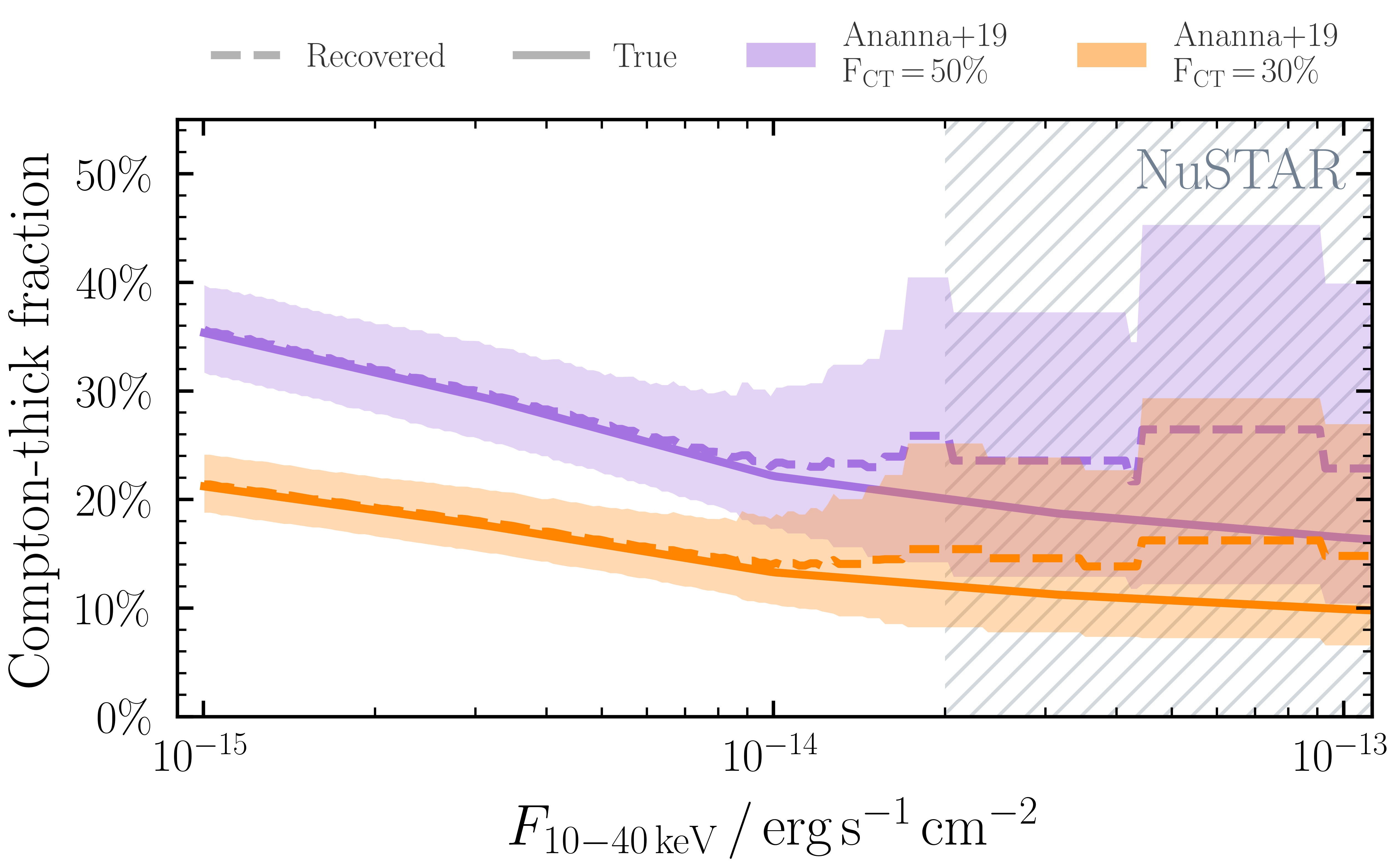}

\caption{\label{ctfrac} The ability to discern population synthesis model predictions of the CT fraction as a function of flux limit in the 10--40 keV band with the simulated wide and deep \textit{HEX-P} surveys combined. The true predicted CT fractions from the population synthesis models of \citet{ananna19} are shown with solid lines for 30\% and 50\% in orange and purple, respectively. The CT fraction recovered purely from CT sources detected above each flux limit on the x-axis is shown with dashed lines and associated binomial uncertainty shading. Clearly the two models can be differentiated by reaching fainter flux levels than \textit{NuSTAR} is currently able to (hatched shading).}
\end{figure}

Constraining the space densities of CT objects has remained elusive due to a lack of high-energy X-ray telescopes capable of detecting faint sources. \textit{HEX-P} can help fill this gap in several ways. Combining spectral analysis of the brightest sources with hardness and band ratio analyses, it will be possible to place constraints on the CT fraction at fluxes ten times, or more, fainter than has been reported before \citep{Burlon11,ricci15,civano15,masini18,torres21}. To visualise the importance of flux limits in assessing the CT fraction, we show the contribution to the CXB from CT objects using the \citet{ananna19} model with two different assumed CT fractions (30\% and 50\%) as solid lines in Figure \ref{ctfrac}. For a series of increasing flux depths, we then considered the detected CT sources above that flux level and re-sampled a random number of not-CT AGN detected above that flux level to give a fraction as close as possible to the value given by each model. We note that since the simulations performed in this work follow a mock catalog based on \cite{gilli07}, we are not directly comparing the measured CT fraction in simulations with the \citet{ananna19} models. Still, we derive uncertainties on the curves in Figure \ref{ctfrac} based on the binomial uncertainties associated with the number of CT sources detected in the surveys above a given flux. For the comparison in Figure \ref{ctfrac}, the sources detected in the 10-40 keV band in the combined deep and wide surveys were considered.

At the flux level sampled by \nustar\ surveys ($>$2$\times$10$^{-14}$ \cgs\ in the 10--40 keV band), the models are indistinguishable. At deeper flux levels, the number of CT sources detected by \textit{HEX-P} surveys are sufficient to constrain the CT fraction to a flux limit of $\sim 10^{-15}$ erg s$^{-1}$ cm$^{-2}$ in the 10--40 keV and inform population synthesis models.

There are several other ways in which \textit{HEX-P} will help us constrain CT space densities. One important approach would be to consider the entire population detected by \textit{HEX-P} and infer the CT fraction with multi-wavelength diagnostics and machine learning algorithms, in particular using infrared colors \citep{stern05,stern12,Mateos13,assef15,Carroll21,carroll23, silver23}. Moreover, given the improved PSF of \textit{HEX-P} compared to \nustar, it will be possible to perform stacking analysis \citep[see the results in][]{Hickox24} at the position of non-X-ray detected candidate CT AGN selected from multiwavelength analysis. We can then apply the \textit{HEX-P} wide and deep survey flux limits in the 10--40 keV energy window to population synthesis models and compare the results with the \textit{HEX-P} observed sample to find the most accurate models. This will be a useful check not only for CT sources but also for relatively unobscured sources, as there are significant disagreements between population synthesis models at Compton-thin obscuration levels as well (as these models have to add up to reproduce the CXB, overestimation in one obscuration level needs to be compensated for by reducing the contribution in others, and vice versa). 
Overall, \textit{HEX-P}'s data promises a deeper, more nuanced insight into AGN space densities, shedding light on the evolution of AGN across cosmic time.

\section{Summary} 
In this paper, we have presented a small selection of the results that will be obtained by performing wide and deep extragalactic surveys with the \textit{High Energy X-ray Probe}. These findings are based on extensive simulations using AGN mock catalogs derived from \cite{marchesi20}. We followed the standard survey detection process to analyze these simulations, leading to the detection of several thousand sources above 3 keV, in contrast to the few hundred sources detected in all of \nustar's surveys performed to date. Notably, we have also detected the first sample of AGN in the 35--55 keV energy band. 

The analysis of these simulations demonstrates that \textit{HEX-P} will, for the first time,  have the capability to significantly constrain the contribution of AGN to the CXB around its peak (3--40 keV). We anticipate reaching a resolved fraction of approximately $\sim$86\% in the 3--40 keV energy range, comparable to what \chandra\ has resolved in the 0.5--8 keV range. This result was achieved using HET detections alone, and even higher resolved fractions can be obtained by performing a stacking analysis of the HET data at the position of LET sources. Furthermore, the resolved fraction could be computed in the 0.5--15 keV energy band exclusively from LET detections. 

With the large samples of sources \textit{HEX-P} will detect, it will be possible to compute the resolved fraction by separating sources in different classes based on X-ray (obscured, unobscured, CT or C-Thin) or optical classifications, allowing to study their relative contribution (and spectrum) at different energies to the total CXB with the ultimate goal of informing population synthesis models and understanding the full cosmic history of black hole growth. 

Understanding the spectrum of the AGN population is a crucial aspect when calibrating population synthesis models. \textit{HEX-P}'s simultaneous broadband coverage will enable spectral analysis for $\sim$1000 sources (combining the LET and HET detections) with sufficient counts to measure and strongly constrain spectral parameters beside obscuration and spectral slope, including, e.g., the covering factor and high-energy cut-off. For sources with low-count spectra, we have shown that the combination of soft and hard X-ray colors can estimate their obscuration level by comparing these colors with the simulations presented in this paper and future simulations. For the first time, we will measure the CT fraction in the 10--40 keV band at a flux limit that is one order of magnitude fainter than \nustar\ with a sample of CT AGN large enough to finally calibrate the next generation of population synthesis models.

When comparing the uncertainty on the measured CXB resolved fraction with the typical difference between AGN population synthesis models, it becomes evident that these new \textit{HEX-P} measurements will be able to disentangle models. It is also plausible that no currently existing model will offer a perfect fit to the population that \textit{HEX-P} will uncover. Together with dedicated observations of local CT AGN \citep{boorman23}, the survey data will open the door to a recalibration of existing population synthesis models, facilitating the integration of more precise constraints on both space densities and AGN spectra.

The potential residing in \textit{HEX-P} surveys extends far beyond what is covered in this paper. For example, we have not discussed the great potential of serendipitous detections of AGN pairs in surveys \citep[see discussion in][]{Pfeifle23} nor the measurements of SMBH spins for the brighter AGN \citep{Piotrowska2023} and also the possible detection of blazars in the wide area survey \citep{marcotulli23}. By the time \textit{HEX-P} is launched, comprehensive spectroscopic campaigns and photometric data will be available for the majority of the extragalactic fields that \textit{HEX-P} will target from major ground and space-based observatories. These data will allow the complete characterization of the multiwavelength properties of \textit{HEX-P} detected AGN as well as the properties of their host galaxies. Understanding the connection between SMBH growth and host galaxy evolution is one of the main goals of the Astro2020 Decadal Survey \citep{Astro_2020} and the \textit{HEX-P} survey samples, combined with these exquisite and rich multiwavelength datasets, will allow us to finally address this goal in the early 2030s.


\section*{Conflict of Interest Statement}

The authors declare that the research was conducted in the absence of any commercial or financial relationships that could be construed as a potential conflict of interest.

\section*{Author Contributions}
This paper is the result of a year of intense analysis and bi-weekly fruitful conversations among the co-authors of this paper. FC is the principal editor of the manuscript and provided guidance and leadership on the entire work. XZ performed the entire analysis of the simulations and measured the resolved fraction. PGB provided significant inputs on the analysis of the CT fraction. SM and C-TC performed the simulations using the \sixte\ code. SC led the spectral analysis of the obscured sources. TA provided significant inputs on the text and on the analysis relevant to population synthesis models. RS contributed to the hardness-ratio analysis. All authors contributed to the final editing of the manuscript.



\section*{Acknowledgments}
SC and RS acknowledges support from the \nustar\ grant number 19-NuSTAR-0036.
CR acknowledges support from the Fondecyt Regular grant 1230345 and ANID BASAL project FB210003. 



\bibliographystyle{Frontiers-Harvard} 

\bibliography{HEX-P_surveys.bib}


\end{document}